\documentclass[a4paper,11pt]{article}
\pdfoutput=1

\usepackage[utf8]{inputenc}
\usepackage{jheppub}
\hypersetup{pdfencoding=unicode, bookmarksopen=true, bookmarksnumbered}
\usepackage[T1]{fontenc}
\usepackage{microtype}
\usepackage{mathrsfs}
\usepackage{subcaption}
\usepackage[all]{xy}
\usepackage{tikz}
\usepackage{diagbox}
\usepackage{physics}
\usepackage{longtable}
\usepackage{afterpage}
\usepackage{lscape}

\usetikzlibrary{decorations.markings}
\usetikzlibrary{arrows.meta}
\tikzset{->-/.style={decoration={
  markings,
  mark=at position #1 with {\arrow{>}}},postaction={decorate}}}

\newcolumntype{L}[1]{>{\raggedright\let\newline\\\arraybackslash\hspace{0pt}}m{#1}}
\newcolumntype{C}[1]{>{\centering\let\newline\\\arraybackslash\hspace{0pt}}m{#1}}
\newcolumntype{R}[1]{>{\raggedleft\let\newline\\\arraybackslash\hspace{0pt}}m{#1}}

\pgfarrowsdeclare{:}{:}{}{}

\let\a=\alpha \let\b=\beta \let\g=\gamma \let\d=\delta \let\e=\epsilon
\let\z=\zeta  \let\th=\theta  
\let\l=\lambda \let\m=\mu \let\n=\nu   \let\r=\rho
\let\s=\sigma     
   \let\G=\Gamma \let\D=\Delta  \let\L=\Lambda
  \let\S=\Sigma

\DeclareMathOperator{\Li}{Li}

\def\a{\alpha}
\def\b{\beta}

\def\CA{{\cal A}}
\def\CB{{\cal B}}
\def\CC{{\cal C}}

\def\CF{{\cal F}}

\def\CH{{\cal H}}
\def\CI{{\cal I}}

\def\CL{{\cal L}}
\def\CM{{\cal M}}
\def\CN{{\cal N}}
\def\CO{{\cal O}}

\def\CS{{\cal S}}

\def\CV{{\cal V}}
\def\CW{{\cal W}}

\def\CZ{{\cal Z}}

\newcommand{\ang}[1]{\langle #1 \rangle}

\def\beq#1\eeq{\begin{align}#1\end{align}}

\makeatletter
\newcommand*{\rom}[1]{\expandafter\romannumeral #1}

\makeatother%For Roman Numeral

\title{The Gaiotto–Kim Trace Formula and Line Defects}

\abstract{
We study the insertion of 4d half-BPS line defects into the trace formula proposed by Gaiotto and Kim, which computes the ellipsoid partition function of the 3d topological quantum field theory (TQFT) arising from a 4d $\mathcal{N}=2$ superconformal field theory. This yields a homomorphism to 3d Wilson loops, preserving the fusion algebra of the 4d defects. We also present a systematic algorithm for determining the simple objects of a 3d TQFT in terms of Wilson loops in the UV 3d $\mathcal{N}=2$ abelian Chern–Simons-matter theory. Utilizing supersymmetric localization, we characterize the modular $S$- and $T$-matrices of the TQFT. We apply our algorithm to various 3d TQFTs arising from the trace formula with higher-power monodromy of the $(A_1,G)$ Argyres-Douglas theories.
}

\author[a]{Sungjoon Kim,}
\author[b]{Spencer Stubbs}

\affiliation[a]{Department of Physics, Korea Advanced Institute of Science and Technology, Daejeon 34141, Korea}
\affiliation[b]{NHETC and Department of Physics and Astronomy, Rutgers University, 126 Frelinghuysen Rd.,
Piscataway NJ 08855, USA}

\emailAdd{sungjoon@kaist.ac.kr}
\emailAdd{spencer.stubbs@physics.rutgers.edu}

\begin{document} 
\maketitle
\flushbottom

%%%%%%%%%%%%%%%%%%%%%%%%%%%%%%%%%%%%%%%%%%%%%%%%%%%%%%%%%%%%%%%%%%%%%%%%
%%%%%%%%%%%%%%%%%%%%%%%%%%%%%%%%%%%%%%%%%%%%%%%%%%%%%%%%%%%%%%%%%%%%%%%%

%%%%%%%%%%%%%%%%%%%%%%%%%%%%%%%%%%%%%%%%%%%%%%%%%%%%%%%%%%%%%%%%%%%%%%%%

\section{Introduction}
Quantum field theories (QFTs) often exhibit rich and unexpected algebraic structures. One notable instance is the map from 4d $\CN=2$ superconformal field theories (SCFT) to vertex operator algebras (VOA), known as the SCFT/VOA correspondence~\cite{Beem:2013sza}, which has inspired extensive study~\cite{Beem:2014rza,Cecotti:2015lab,Lemos:2015orc,Xie:2016evu,Song:2016yfd,Song:2017oew,Fluder:2017oxm,Fredrickson:2017yka,Beem:2017ooy,Beem:2019tfp}; see also~\cite{Kim:2024dxu,Kim:2025klh,Kang:2025zub,Kang:2026nge,Maruyoshi:2026cmr} for recent developments. The statement is that a sub-sector of a SCFT, known as the Schur sector, captures a certain VOA structure. Further evidence for this is provided by a wall-crossing formula for the Schur index based on the BPS particle spectrum on the Coulomb branch, yielding the vacuum character of the associated VOA~\cite{Cordova:2015nma}. Half-BPS line defects are naturally coupled to this structure. In particular, a generating function of the UV line defect into IR quantum torus algebra variables allows one to define a defect version of the Schur index~\cite{Cordova:2016uwk}. The Schur index with a line defect insertion computes a superposition of the VOA characters. Moreover, it preserves the fusion rules of the defects, defining a homomorphism to a Verlinde-like algebra~\cite{Neitzke:2017cxz}.

A different wall-crossing invariant, defined as the trace of a monodromy operator $\Tr \Phi$ over an auxiliary Hilbert space has recently been proposed~\cite{Gaiotto:2024ioj}. It computes the ellipsoid partition function of a 3d $\CN=2$ abelian Chern-Simons matter (ACSM) theory arising from the $U(1)_r$ twisted circle reduction of a 4d $\CN=2$ SCFT~\cite{Go:2025ixu,Kim:2025rog,Nishinaka:2025ytu}. Surprisingly, the resulting theory flows to a topological quantum field theory (TQFT) upon 3d topologcial twist and supports the expected VOA on its holomorphic boundary~\cite{Gaiotto:2024ioj,ArabiArdehali:2024ysy,ArabiArdehali:2024vli,Go:2025ixu,Kim:2025rog,Nishinaka:2025ytu,Closset:2026xjj,Yoshida:2026wvb,Kim:2026ghk}. Stated differently, this wall-crossing formula provides a 3d TQFT that mediates the SCFT/VOA correspondence. As a natural extension, higher powers of the monodromy trace $\Tr \Phi^n$ --- corresponding to $n$-th wrapping $U(1)_r$ twisted circle reduction --- can be considered \cite{Go:2025ixu}. The resulting TQFTs form a finite orbit, with their modular data related by Galois transformations.

In this paper, we explore the UV line defect insertion in the proposed trace formula. By replacing the quantum torus algebra variables with those of a Weyl algebra, we find that the defects are mapped to Wilson loops of the 3d $\CN=2$ ACSM theory. This is analogous to the homomorphism discussed in~\cite{Neitzke:2017cxz} that preserves the fusion algebra of the line defects. We also show explicitly for the $(A_1,A_2)$ and $(A_1,A_3)$ Argyres-Douglas (AD) theories, that this homomorphism itself is preserved under increasing the monodromy power. In what follows, we will use the variable component $G$ as shorthand for the $(A_1,G)$ AD theory.

Additionally, we present a systematic algorithm for determining a set of Wilson loops in a 3d $\CN=2$ UV ACSM theory that correspond to the simple objects, or anyons, of the IR 3d TQFT. This utilizes a set of necessary conditions from the twisted partition function and superconformal index computations, and reconstructs the full modular $S$- and $T$-matrices, characterizing the anyon dynamics of the TQFT. We apply this algorithm to the TQFTs obtained from higher monodromy traces, considering all $G$ theories with $\text{rank}(G)\leq 6$, thereby determining the associated modular data. Based on the classification in~\cite{ng2025classificationmodulardatarank}, we confirm a set of modular data from higher monodromy trace for every $G$ forms a Galois orbit.

This paper is organized as follows. We begin by reviewing the pioneering works on the wall-crossing formula in section \ref{sec: 1.1}. Then, in section \ref{sec: 2}, we review the proposed wall-crossing formula $\Tr \Phi$ and discuss the UV line defect insertion, exhibiting the fusion-preserving homomorphism in terms of Wilson loops. In section \ref{sec: 3}, we present an algorithm for determining simple lines in 3d TQFT as Wilson loops in the UV ACSM theory and provide various examples from the trace formula of the $G$ theories up to $\text{rank}(G)\leq 6$. In appendix \ref{app: A-model}, we summarize supersymmetric localization formulae, with particular emphasis on ACSM theories. In appendix \ref{app: A3 monodromy action}, we provide supplementary computations regarding the $A_3$ theory. 

\subsection{The IR formula for the Schur index and line defects}\label{sec: 1.1}
The Schur index $\CI(q)$ of a 4d $\CN=2$ SQFT can be evaluated in the IR electrodynamic description on the Coulomb branch~\cite{Cordova:2015nma}
\begin{align}
    \CI(q) = (q)_{\infty}^{2r}\Tr[ \CO(q) ]
    \,,
\end{align}
where $(q)_\infty$ is the $q$-Pochhammer symbol, $r$ is the dimension of the Coulomb branch, and $\CO(q)$ is the quantum monodromy operator. This is defined as an ordered product over the BPS spectrum of charge $\g$ arranged in increasing order of their central charge phases
\begin{align}
    \CO(q) = 
    \prod_{\g }^{\curvearrowleft}
    E_q(X_\g)
    \;\; ;\;\;\;
    E_q(z) 
    = \prod_{i=0}^\infty(1+z q^{i+\frac{1}{2}})^{-1}
    \,,
\end{align}
where $X_\g$ is the quantum torus algebra variable. The $X_\g$ have a natural interpretation as an abelian 't Hooft-Wilson line defects and $q$ measures their non-commutativity through the electromagnetic angular momentum~\cite{Cordova:2016uwk}
\begin{align}
    X_\g X_{\g'} = q^{\frac{1}{2}{\langle \g,\g' \rangle}} X_{\g+\g'}
    \,.
    \label{eq: qtalg commt}
\end{align}

\medskip
\noindent{\bf Line defect Schur index.}
A half-BPS line defect $L$ defined in the UV is mapped to a superposition of the abelian defects $X_\g$ in the IR which is encoded by a generating function~\cite{Gaiotto:2010be,Cordova:2013bza}
\begin{align}
    F(L,\th) = \sum_\g \underline{\overline{\Omega}}(L,\th,\g,q) X_\g
    \,,
\end{align}
where $\th$ is the insertion angle of $L$ along the half-BPS great circle of $S^3$, and $\underline{\overline{\Omega}}(L,\th,\g,q)$ is the framed BPS degeneracy. The Schur index modified by a line defect $L$ is given by~\cite{Cordova:2016uwk}
\begin{align}
    \CI_L(q) = (q)_\infty^{2r} \Tr[ F(L) \CO(q) ]
    \,,
\end{align}
where we suppressed $\th$ dependence by placing $F(L,\th)$ to the right of all ordinary BPS particles since the angle coincides with the central charge phase of the BPS state. This modification respects the defect OPE with coefficients $c_{\a\b}^\r \in \mathbb{Z}_{\geq 0}[q^{\frac{1}{2}},q^{-\frac{1}{2}}]$
\begin{align}
    L_\a L_\b = 
    \sum_{\r} c_{\a\b}^\r(q) L_\r
    \;\;\to\;\;
    \CI_{L_\a L_\b}(q) = 
    \sum_{\r} c_{\a\b}^\r(q) \CI_{L_\r}(q)
    \,,
    \label{eq: L OPE}
\end{align}
which holds because the OPE is invariant under RG flow due to supersymmetry.

\medskip
\noindent{\bf Relation to VOA.}
Regarding the SCFT/VOA correspondence~\cite{Beem:2013sza}, the line defect Schur index of a 4d $\CN=2$ SCFT computes a superposition of the corresponding VOA characters~\cite{Cordova:2016uwk}
\begin{align}
    \CI_{L_{\a_i}}(q) = \sum_{\r} v_{\a_i}^{\r}(q) \chi_\r (q)
    \,,
\end{align}
where the subscript $i$ of $\a$ labels a possible degeneracy for a given $\a$, and the coefficients $v_{\a_i}^\r(q)$ are polynomials in $q$. Likewise, we write the index for a defect fusion as
\begin{align}
    \CI_{L_{\a_i}L_{\b_j}}
    =
    \sum_{\r} v_{\a_i \, \b_j}^\r(q) \chi_\r(q)
    \,.
\end{align}
The coefficients become integers as $q \to 1$, losing their degeneracy dependence
\begin{align}
    V_{\a}^{\r} := v_{\a_i}^\r (q=1)
    \;,\;\;
    V_{\a\b}^{\r} := v_{\a_i\b_j}^\r (q=1)
    \,,
\end{align}
which satisfy the Verlinde algebra for the fusion coefficient $N_{\r\s}^{\l}$ of the associated VOA
\begin{align}
    V_{\a\b}^\l
    =
    \sum_{\r,\s} N_{\r\s}^{\l} V_\a^\r V_\b^\s
    \,.
    \label{eq: VA}
\end{align}
Then, by writing the Verlinde algebra elements with basis $[\![\varphi_\r]\!]$ corresponding to chiral primaries of the VOA as
\begin{align}
    [\![L_\a]\!] = \sum_\r V_\a^\r [\![\varphi_\r]\!]
    \quad,\quad
    [\![L_\a L_\b]\!] = \sum_\r V_{\a\b}^\r [\![\varphi_\r]\!]
    \,,
\end{align}
the property \eqref{eq: VA} implies the existence of a fusion-preserving homomorphism
\begin{align}
    [\![L_\a L_\b]\!] = [\![L_\a]\!] \times [\![L_\b]\!]
    \,.
    \label{eq: L fusion}
\end{align}
In the remainder of the paper, we examine these aspects of defects also hold for a recently proposed alternate trace~\cite{Gaiotto:2024ioj}, which we introduce in the next section.

\section{Trace formula and line operator insertion} \label{sec: 2}
In this section we review the trace formula introduced in~\cite{Gaiotto:2024ioj} and discuss the insertion of a half-BPS UV line defect into it. This yields a natural homomorphism to 3d supersymmetric Wilson loops in the 3d $\CN=2$ abelian gauge theory arising from the trace formula. This preserves the fusion algebra regardless of the powers of the monodromy in the trace.

\subsection{Trace formula}
The Gaiotto--Kim trace formula computes the ellipsoid partition function of a 3d $\CN=2$ abelian gauge theory arising from $U(1)_r$ twisted circle reduction of a 4d $\CN=2$ SCFT~\cite{Gaiotto:2024ioj}
\begin{align}
    Z_{S_b^3} = 
    \Tr_{\CH}
    \Phi
    \, ,
    \label{eq: trace formula}
\end{align}
where the trace is over an auxiliary Hilbert space $\CH = L^2(\mathbb{R}^{\rank(\G/\G_f)/2})$ with charge lattice $\G$ of the 4d BPS particle charges $\g$, and $\G_f$ being its sub-lattice for the Higgs branch flavor symmetry. The monodromy operator $ \Phi$ is given by
\begin{align}
     &\qquad\qquad 
     \Phi
     =
    \prod_{\g \in \G_{\text{BPS}}}^{\curvearrowleft} \Phi_b(x_\g)
    \prod_{\g \in \G_{\text{BPS}}}^{\curvearrowleft} \Phi_b(x_{-\g})
     \,,
     \nonumber\\
     &\text{with}\;\;
     \Phi_b(x) = \exp\left( \frac{1}{4} \int_{\mathbb{R}+i 0^+} \frac{dt}{t}\frac{e^{-2i\, x \, t}}{\sinh(b\,t)\sinh(b^{-1}\,t)} \right)
     \,,
    \label{eq: monodromy}
\end{align}
where the two ordered products of the Faddeev's quantum dilogarithms (QDLs)~\cite{Faddeev:1993rs} in $ \Phi$ are over the charge spectrum $\G_\text{BPS}$ of BPS particles, arranged in increasing order of their central charges, $\CZ_\g : \G \to \mathbb{C}$, which is a complex valued linear function of $\g$. The variables $x_\g$ obey the Weyl algebra determined by the Dirac paring
\begin{align}
    [x_\g , x_{\g'}] = \frac{1}{2\pi i} \langle \g,\g' \rangle \,.
\end{align}
The QDL follows {\it pentagon} identity if $\langle \g,\g' \rangle=1$ that encapsulates the wall-crossing property
\begin{align}
    \Phi_b(x_\g) \Phi_b(x_{\g'})
    =
    \Phi_b(x_{\g'})
    \Phi_b(x_\g+x_{\g'})
    \Phi_b(x_\g) ,
    \label{eq: pentagon id}
\end{align}
and together with the fusion identity
\begin{align}
    \Phi_b(x) \Phi_b(-x) = e_b^2\, e^{\pi i \, x^2}
    \;\; ; \;\;\;
    e_b := e^{\frac{\pi i}{24}(b^2 + \frac{1}{b^2}) }
    \,,
    \label{eq: fusion id}
\end{align}
the trace can be simplified. See \cite{Kim:2025rog} for this simplification using {\tt Mathematica}. We emphasize that the trace \eqref{eq: trace formula} is a wall-crossing invariant, so the result is independent of the choice of chamber on the Coulomb branch.

\medskip
\noindent {\bf Example: \texorpdfstring{$A_2$}{A2} theory.}
As a quick illustration, consider the $A_2$ theory whose minimal chamber contains two BPS particles with the ordered charge set $\G_{\text{BPS}} = \{\g_1,\g_2\}$, and the Dirac pairing $\langle \g_1,\g_2 \rangle = 1$:
\begin{align}
    Z_{S_b^3}^{A_2} &= \Tr \big(
    \Phi_b(x_{\g_1}) \Phi_b(x_{\g_2}) \Phi_b(-x_{\g_1}) \Phi_b(-x_{\g_2})
    \big)
    \nonumber\\
    & = 
    e_b^4
    \Tr \big(
    e^{\pi i x_{\g_2}^2} 
    \Phi_b(x_{\g_1}+x_{\g_2})
    e^{\pi i x_{\g_1}^2} 
    \big)
    \nonumber\\
    & = 
    e_b^4
    \Tr \big(
    e^{\pi i x_{\g_2}^2} 
    e^{\pi i x_{\g_1}^2} 
    \Phi_b(x_{\g_2})
    \big)
    \,,
\end{align}
where in the second line we used the pentagon and fusion identities, then a shift in the third line:
\begin{align}
    e^{\pi i x_{\g_1}^2} f(x_{\g_2}) = f(x_{\g_1}+x_{\g_2}) e^{\pi ix_{\g_1}^2}
    \,.
\end{align}
The Weyl algebra variables can be regarded as momentum and position operators in an auxiliary quantum mechanics as $x_{\g_1} = \hat{p}$ and $x_{\g_2} = \hat{x}$ so that we can adopt the below tools
\begin{align}
    &\;\;\;
    [\hat{p},\hat{x}] = \frac{1}{2\pi i}\;,\;\;
    \hat{p} \ket{p} = p \ket{p}
    \;,\;\;
    \hat{x} \ket{x} = x \ket{x}\,,
    \nonumber\\
    &\langle x|p \rangle = e^{2\pi i \, p x} 
    \;,\;\;
    {\bf 1} = \int dp \, |p \rangle \langle p|
    = \int dx \, |x \rangle \langle x|
    \,,
\end{align}
to evaluate
\begin{align}
    Z_{S_b^3}^{A_2} &=i^{\frac{1}{2}} e_b^4
    \int d\s \, e^{\pi i \s^2} \Phi_b(\s)
    \,.
    \label{eq: GY int}
\end{align}
Such an embedding of the Weyl algebra to quantum mechanics is always possible; see the appendix of \cite{Kim:2025rog} for more details and examples.

The integral expression \eqref{eq: GY int} coincides with the ellipsoid partition function of the Gang--Yamazaki minimal rank-0 SCFT \cite{Gang:2018huc} in the A-twisted sector, capturing a 3d TQFT with the modular structure of the $M(2,5)$ minimal model \cite{Gang:2023rei}. This precisely coincides with the associated VOA of the $A_2$ theory via the SCFT/VOA correspondence \cite{Beem:2013sza,Beem:2014cca,Beem:2014rka,Cordova:2015nma,Beem:2017ooy}. 

This 3d TQFT, admitting the boundary VOA associated to the 4d SCFT, was explained in \cite{Dedushenko:2023cvd} through the $U(1)_r$ twisted cigar circle reduction of the SCFT, following the holomorphic-topological twist \cite{Kapustin:2006hi} with an $\Omega$-deformation along the cigar \cite{Oh:2019bgz,Jeong:2019pzg}. Indeed, for various $(G,G')$ AD theories, it was verified in~\cite{Go:2025ixu,Kim:2025rog} that the trace \eqref{eq: trace formula} encodes an intermediate 3d TQFT between the 4d SCFT and the 2d VOA.\footnote{See \cite{ArabiArdehali:2024ysy,ArabiArdehali:2024vli}, for an alternative approach based on the high temperature effective theory of the 4d $\CN=1$ Maruyoshi--Song's Lagrangian descriptions of the AD theories \cite{Maruyoshi:2016tqk,Maruyoshi:2016aim,Agarwal:2016pjo}. 
} Namely, one can investigate the SCFT/VOA correspondence using the trace \eqref{eq: trace formula}, with the intermediate 3d TQFT in hand.

\subsection{Trace formula with lines \label{sec: 4d to 3d}}
Now, let us consider the line defect insertion in the trace formula. Note that the quantum torus algebra variable $X_\g$ is mapped to the Weyl algebra variable $x_\g$ as
\begin{align}
    X_\g = e^{2\pi b^{\pm 1} x_\g}
    \,,
    \label{eq: half BPS 3d defect}
\end{align}
which preserves the non-commutativity in \eqref{eq: qtalg commt} for $q = e^{-2\pi i b^{\pm 2}}$. We note that the operator in \eqref{eq: half BPS 3d defect} corresponds to the 3d supersymmetric Wilson, or 't Hooft loop operator studied in \cite{Beem:2012mb,Dimofte:2011jd,Dimofte:2011py}. Indeed, as the trace is evaluated, the eigenvalue of $x_\g$ becomes the Coulomb parameter in the ellipsoid partition function expression and  \eqref{eq: half BPS 3d defect} contributes as a half-BPS Wilson loop the 3d $\CN=2$ abelian gauge theory. We will continue to use the notation $L$ for the line defect in the trace formula \eqref{eq: trace formula}, with the change of variables \eqref{eq: half BPS 3d defect} understood. 

On the squashed three-sphere $S_b^3$, the half-BPS Wilson loop $W$ is closed only when it is supported at either the north or south pole, which are the cores of the two $D^2\times S^1$ solid tori in the Hopf decomposition. It appears in the partition function as an exponential factor
\begin{align}
    Z_{S_b^3}\big(W\big)
    =
    \int d\s\, e^{2\pi b^{\pm 1}Q \s}
    (\cdots)
    \,,
\end{align}
where $Q$ is the charge of the loop $W$, while $(\cdots)$ is determined by the details of the gauge theory, and $b$/$b^{-1}$ in the exponent corresponds to the north/south pole insertion of $W$ respectively. In the remainder of the section, we focus on the north pole insertion as the analysis for the south pole is completely analogous upon replacing $b$ with $b^{-1}$.

In general, the trace with the insertion of $F(L)$ gives rise to a superposition of the 3d Wilson loops in the 3d $\CN=2$ abelian gauge theory obtained by $Z_{S_b^3} = \Tr(\Phi)$
\begin{align}
    \Tr\big(F(L) \Phi\big) = \sum_{W} c_W^L(q) Z_{S_b^3}(W)
    \,,
    \label{eq: FL to ZW}
\end{align}
with polynomials $c_W^{L}(q) \in \mathbb{Z}[q^{1/2},q^{-1/2}]$. For the given 3d gauge theory, let us define
\begin{align}
    \langle L \rangle_b := \Tr\big(F(L) \Phi\big)
    \;,\;\;
    \langle W \rangle_b := Z_{S_b^3}(W)
    \,,
\end{align}
so that \eqref{eq: FL to ZW} is rewritten compactly as
\begin{align}
    \langle L \rangle_b
    =
    \sum_{W} c_W^L(q) 
    \langle W \rangle_b
    \,.
    \label{eq: L to W}
\end{align}

\medskip
In the remainder of this section, we focus on two examples; $A_2$ and $A_3$ theories. We first show the action of $\Phi$ on $F(L)$ in analogy with the quantum monodromy operator~\cite{Cordova:2016uwk,Neitzke:2017cxz}
\begin{align}
    \Phi F(L_{\a_{i}})  =  F(L_{\a_{j}}) \Phi
    \,,
    \label{eq: monodromy action}
\end{align}
for $i\neq j$ which labels the degeneracy of $L_\a$. This ensures the $i$-independence upon taking the trace, hence, we define the ellipsoid partition function of $L_\a$ insertion without $i$ as
\begin{align}
    \langle L_\a \rangle_b
    :=
    \langle L_{\a_i} \rangle_b
    \,.
\end{align}
Next, we determine the Wilson lines $W$ and their coefficients $c_W^L(q)$ for a given $L$. We employ the Bethe root method~\cite{Nekrasov:2014xaa,Benini:2015noa,Benini:2016hjo,Closset:2016arn} for the resulting 3d $\CN=2$ gauge theory, whose vacuum equation encodes the fusion rule of the $W$ at $b=1$. Equipped with this, we recover the OPE of $L$ with the map \eqref{eq: L to W}. We also consider higher-power monodromy trace with $L$ insertion:
\begin{align}
    \ang{L}_b^{(n)}
    :=
    \Tr ( F(L) \Phi^n )
    \,,
\end{align}
for $n>1$. By using the 3d A-model method, we confirm the fusion-preserving homomorphism
\begin{align}
    \ang{L L'}_{b=1}^{(n)}
    =
    \ang{L}_{b=1}^{(n)}
    \times
    \ang{L'}_{b=1}^{(n)}
\end{align}
which extends the notion of the homomorphism at $n=1$~\cite{Cordova:2016uwk,Neitzke:2017cxz} to general $n$.

\medskip
\noindent {\bf Abbreviations.}
For notational simplicity, we use some abbreviations through the examples:
\begin{align}
    X_a := X_{\g_a}
    \;\;,\;\;\;\;\;
    \CA_i, \CB_i, \CC_i,... :=  L_{\a_i}
    \,,
\end{align}
where we label different species of line defects by curly letters, only leaving the degeneracy index. Also, we use
\begin{align}
    (a) := \Phi_b(x_{\g_a})
    \;\;,\;\;\;
    e^a := 
    e_b^{2}
    e^{\pi i x_{\g_a}^2}
    \;\;,\;\;\;
    [a] := e^{2\pi b x_{\g_a}}
    \,,
    \label{eq: abbreviation}
\end{align}
for the relevant operators in the trace formula evaluation. Then, we rewrite the shift properties by the last two as (for $\langle \g_a , \g_{a'} \rangle = k$)
\begin{align}
    % e^{\pi i x_{\g_a}^2} f(x_{\g_{a'}}) 
    % = f(x_{\g_{a'}}+ k x_{\g_a}) e^{\pi i x_{\g_a}^2}
    % \;&\overset{f=\Phi_b}{\to}\;\;\;\;\;
    e^a f(a') = f(a' + k\cdot a) e^a
    \;\;,\;\;\;\;
    % e^{2\pi b x_{\g_a}} f(x_{\g_{a'}}) 
    % = f(x_{\g_{a'}} -i k b) e^{2\pi b x_{\g_a}}
    % \;&\overset{f=\Phi_b}{\to}\;\;\;\;\;
    [a] f(a') = f(a' - i k b) [a]
    \,,
    \label{eq: shift properties}
\end{align}
for some function $f(x_{\g_{a'}}):=f(a')$ of $x_{\g_{a'}}$. We use a dot $\cdot$ before the Weyl variable to distinguish it from ordinary multiplication. We also write the quasi-periodicity of the QDL
\begin{align}
    % \Phi_b(x_{\g_a}\pm ib) = \big\{1+e^{\pm \pi i b^2} e^{2\pi b x_{\g_a}} \big\}^{\mp 1} \Phi_b(x_{\g_a})
    % \;\to\;\;\;\;
    (a \pm i b) = \big\{ 1+ q^{\mp \frac{1}{2}} [a] \big\}^{\mp 1}
    (a)
    \,.
\end{align}

\subsubsection{\texorpdfstring{$A_2$}{A2} theory}
Let us first consider the $A_2$ theory which has one non-trivial half-BPS defect with five-fold degeneracy~\cite{Cordova:2016uwk}:
\begin{align}
    &F(\CL_1) = X_1
    \,,
    \nonumber\\
    &F(\CL_2) = X_2 + X_{1+2}
    \,,
    \nonumber\\
    &F(\CL_3) = X_{-1} + X_{-1+2} + X_2
    \,,
    \nonumber\\
    &F(\CL_4) = X_{-1-2} + X_{-1}
    \,,
    \nonumber\\
    &F(\CL_5) = X_{-2}
    \,,
\end{align}
where $X_{i+j+\cdots}$ stands for $X_{\g_i + \g_j + \cdots}$ and similarly elsewhere. These satisfy the defect OPE with the Dirac pairing $\langle \g_1,\g_2 \rangle = 1$ as
\begin{align}
    \CL_i \CL_{i+2} =  {\bf 1} + q^{\frac{1}{2}} \CL_{i+1}
    \,,
    \label{eq: A2 L OPE}
\end{align}
where the indices are periodic mod 5.

\medskip
\noindent {\bf Monodromy action.}
The monodromy operator $\Phi$ can be simplified from the QDL properties \eqref{eq: pentagon id} and \eqref{eq: fusion id} as
\begin{align}
    \Phi = (1)(2)(-1)(-2) = e^1 e^2 (-1)
    \,,
\end{align}
and we check the property \eqref{eq: monodromy action}
\begin{align}
    \Phi F(\CL_5) &= [-1-2] (2-ib) e^1 e^2 
    = \big\{ [-1-2]+[-1] \big\} (2) e^1 e^2 = F(\CL_4) \Phi\,,
    \nonumber\\
    \Phi F(\CL_4) &= \Big\{   
    [-1]\big\{ 1 + q^{\frac{1}{2}}[2] \big\} + [2]
    \Big\}(2) e^1 e^2
    = \big\{ [-1] + [-1+2] + [2] \big\} (2) e^1 e^2 \\
    &= F(\CL_3) \Phi\,,
    \nonumber\\
    \Phi F(\CL_3) &= \Big\{
    [2] + [1+2] \frac{1}{1+q^{-1/2}[2]}
    + q^{-1/2} [1+2][2] \frac{1}{1+q^{-1/2}[2]}
    \Big\} (2) e^1 e^2
    \nonumber\\
    &= \Big\{
    [2]+[1+2]
    \Big\} (2) e^1 e^2 = F(\CL_2) \Phi \,,
    \nonumber\\
    \Phi F(\CL_2) &= 
    \Big\{
    [1][2]q^{-1/2} \frac{1}{1+q^{-1/2}[2]} + [1] \frac{1}{1+q^{-1/2}[2]}
    \Big\} (2) e^1 e^2
    = F(\CL_1) \Phi 
    \nonumber\\
    \Phi F(\CL_1) &= (2) e^1 e^2 [1]
    = [-2] (2) e^1 e^2 
    = F(\CL_5) \Phi\,,
\end{align}
and by the nature of the trace,
\begin{align}
    \langle \CL_{i+1} \rangle_b
    =
     \Tr\big( F(\CL_{i+1}) \Phi \big)
     =
     \Tr\big( \Phi F(\CL_{i})  \big)
     =
     \Tr\big( F(\CL_{i}) \Phi  \big)
     =
     \langle \CL_{i} \rangle_b
    \,,
\end{align}
it verifies the $i$-independence in the trace
\begin{align}
    \ang{\CL_i}_b = \ang{\CL}_b
    \,.
    \label{eq: A2 L indep}
\end{align}

\medskip
\noindent {\bf Map to the Wilson loops.}
A generic monomial of $X_1$ and $X_2$ can be arranged as $X_1^m X_2^n$ for some integers $m,n\in \mathbb{Z}$ up to an overall $q$ factor. Then, its insertion in the trace gives rise to
\begin{align}
    \Tr( X_1^m X_2^n \Phi )
    &= \Tr \big( [m\cdot 1] [n\cdot 2] e^1 e^2 (-1) \big)
    \nonumber\\
    &=e_b^4 \, i^{\frac{1}{2}}
    \int d\s
    \big(e^{2\pi b \s}\big)^{n-m}
    e^{\pi i \s^2}
    \Phi_b(\s)
    \,.
\end{align}
This is the ellipsoid partition function of the GY theory --- obtained from $\Tr(\Phi)$ --- together with a Wilson loop $W_{(n-m)}$ of charge $n-m$ on the north pole. Equipped with this, we evaluate the traces $\langle \CL_i \rangle_b$ as
\begin{align}
    &\langle \CL_1 \rangle_b =
    \langle \CL_5 \rangle_b =
    e_b^4 i^{\frac{1}{2}}
    \int d\s \big(e^{-2\pi b \s}\big) e^{\pi i \s^2} \Phi_b(\s)
    = \ang{W_{(-1)}}_b
    \,,
    \nonumber\\
    &
    \langle \CL_2 \rangle_b =
    \langle \CL_4 \rangle_b =
    e_b^4 i^{\frac{1}{2}}
    \int d\s 
    \big(e^{2\pi b \s} + e^{\pi i b^2} \big)
    e^{\pi i \s^2} \Phi_b(\s)
    = \ang{W_{(1)}}_b + q^{-\frac{1}{2}}\ang{W_{(0)}}_b
    \,,
    \nonumber\\
    &
    \;\;\;\;
    \langle \CL_3 \rangle_b =
    e_b^4 i^{\frac{1}{2}}
    \int d\s 
    \big(2 e^{2\pi b \s} + e^{-\pi i b^2}e^{4\pi b \s} \big)
    e^{\pi i \s^2} \Phi_b(\s)
    = 2\ang{W_{(1)}}_b + q^{\frac{1}{2}}\ang{W_{(2)}}_b
    \,,
\end{align}
where we denote the Wilson loop of charge $Q$ by $W_Q$.

\medskip
\noindent {\bf The line OPE from Bethe root method.}
To make a contact with the Bethe root method as summarized in appendix~\ref{app: A-model}, we set $\s = - i u$ to get the vacuum equation of the GY theory in the A-twisted sector
\begin{align}
    z^2 + z - 1 = 0
    \,,
\end{align}
where $z = e^{2\pi i u}$, thereby at $b=1$ (a round three-sphere), we have a relation $W_{(Q)} = z^{-Q}$. Namely, the vacuum equation, that is a polynomial by construction, encodes the fusion rules of the Wilson loops. We also re-check the degeneracy independence \eqref{eq: A2 L indep} as
\begin{align}
    \ang{\CL_1} &= \ang{\CL_5} = \ang{z}
    \,,
    \nonumber\\
    \ang{\CL_2} &= \ang{\CL_4} = \ang{z^{-1} - 1} = \ang{z}
    \,,
    \quad\;\to\quad
    \ang{\CL} = \ang{z}
    \nonumber\\
    \ang{\CL_3} &= \ang{2 z^{-1} - z^{-2}} = \ang{z}
    \,.
\end{align}
Here, we omitted the subscript $\ang{\cdots}_{b=1}$ to avoid clutter. Furthermore, we find a homomorphism that preserves the fusion algebra \eqref{eq: A2 L OPE} as similarly observed in~\cite{Cordova:2016uwk,Neitzke:2017cxz}
\begin{align}
    \ang{\CL  \CL} 
    =\ang{z^2}
    =\ang{1-z}
    =\ang{\bf{1}} - \ang{\CL}
    =
    \ang{\CL} \times \ang{\CL}
    \,,
    \label{eq: A2 fusion homo}
\end{align}
where $\times$ denotes the fusion algebra product. Note that the minus coefficient is compatible with \eqref{eq: A2 L OPE} as $q^{1/2} = -1$ at $b=1$.

\medskip\noindent{\bf Higher-power monodromy}.
Consider the trace of $n$-th power monodromy with the line insertion
\begin{align}
    \ang{\CL_{i+1}}_b^{(n)}
    =
    \Tr\big( F(\CL_{i+1}) \Phi^n \big)
    =
    \Tr\big( \Phi F(\CL_{i}) \Phi^{n-1} \big)
    =
    \Tr\big( F(\CL_{i}) \Phi^{n} \big)
    =
    \ang{\CL_{i}}_b^{(n)}
    \,,
\end{align}
which shows the degeneracy independence
\begin{align}
    \ang{\CL_i}_b^{(n)}
    =
    \ang{\CL}_b^{(n)}
    \,.
\end{align}
Then, we find the map of $\CL$ to Wilson loops as~\footnote{Here we only consider $n=2,3,4$ since $\Phi^5$ becomes identity operator so that $n>5$ is periodic mod 5.}
\begin{align}
    \ang{\CL}_b^{(2)}
    &=
    i e_b^8 \int d\s_1 d\s_2
    \big( 
    e^{-2\pi b \s_1} 
    \big)
    e^{\pi i ( -\s_1^2 + 4 \s_1\s_2 - \s_2^2 ) }
    \Phi_b(\s_1)\Phi_b(\s_2)
    =
    \ang{W_{(-1,0)}}_b^{(2)}
    \,,
    \nonumber\\
    \ang{\CL}_b^{(3)}
    &=
    i e_b^{12} \int d\s_1 d\s_2
    \big( 
    e^{2\pi b \s_1}
    \big)
    e^{-4\pi i \s_1\s_2}
    \Phi_b(\s_1)\Phi_b(\s_2)
    =
    \ang{W_{(1,0)}}_b^{(3)}
    \,,
    \nonumber\\
    \ang{\CL}_b^{(4)}
    &=
    -i^{\frac{3}{2}} e_b^{18} \int d\s
    \big(
    e^{2\pi b \s}
    \big)
    e^{-2\pi i \s^2} \Phi_b(\s)
    =
    \ang{W_{(1)}}_b^{(4)}
    \,,
\end{align}
which can be written in terms of Bethe equation variables at $b=1$ as
\begin{align}
    &n=2\;:\;
    \ang{\CL}^{(2)}= \ang{z_1}^{(2)}
    \;\;,\;\;
    \frac{z_2^2}{1-z_1} = \frac{z_1^2}{1-z_2} = 1
    \,,
    \nonumber\\
    &n=3\;:\;
    \ang{\CL}^{(3)}= \ang{z_1^{-1}}^{(3)}
    \;\;,\;\;
    \frac{z_1 z_2^{-2}}{z_1-1} = \frac{z_1^{-2}z_2}{z_2-1} = 1
    \,,
    \nonumber\\
    &n=4\;:\;
    \ang{\CL}^{(4)}= \ang{z^{-1}}^{(4)}
    \;\;,\;\;
    \frac{1}{z^2-z} = 1
    \,.
\end{align}
They imply the {\it same} fusion-preserving homomorphism as given in \eqref{eq: A2 fusion homo}
\begin{align}
    \ang{\CL  \CL}^{(n)}
    =
    \ang{\CL}^{(n)}
    \times
    \ang{\CL}^{(n)}
    \,,
\end{align}
for $n=2,3$, and $4$ as well. This result suggests an extended notion of the homomorphism to higher-power monodromy.

\subsubsection{\texorpdfstring{$A_3 = D_3$}{A3=D3} theory}
Consider the $A_3$ theory which has three non-trivial half-BPS line defects $\CA_i$, $\CB_i$, and $\CC$ with degeneracy three~\cite{Cordova:2016uwk}
\begin{align}
    &F(\CA_1) = X_p
    \,,
    \nonumber\\
    &F(\CA_2) = X_{-p-x} + X_{-p}
    \,,
    \nonumber\\
    &F(\CA_3) = X_{-p} + (\xi+\xi^{-1}) X_{x} + X_{p+x} + X_{-p+x}
    \,,
    \nonumber\\
    &F(\CB_1) = X_{-x}
    \,,
    \nonumber\\
    &F(\CB_2) = X_{x} + (\xi+\xi^{-1}) X_{-p} 
    + (\xi+\xi^{-1}) X_{-p+x}
    + (q^{\frac{1}{2}}+q^{-\frac{1}{2}}) X_{-2p}
    +X_{-2p + x} + X_{-2p-x}
    \,,
    \nonumber\\
    &F(\CB_3) = X_x + (\xi+\xi^{-1}) X_{p+x} + X_{2p+x}
    \,,
    \nonumber\\
    &F(\CC)  = \xi+\xi^{-1}
    \,,
\end{align}
where we defined $\xi := e^{2\pi b x_{\g_m}}$ and
\begin{align}
    \g_p := \frac{\g_1 + \g_3}{2}
    \;,\;\;\;
    \g_x := \g_2
    \;,\;\;\;
    \g_m := \frac{\g_1-\g_3}{2}
    \,,
\end{align}
with the only non-trivial Dirac pairing of them is $\ang{\g_p , \g_x} = 1$ and $\g_m$ parametrizes the Higgs branch flavor symmetry. The fusions are then given as
\begin{align}
    &\CA_i \CA_{i+1} = {\bf 1} + q^{-\frac{1}{2}} \CB_i
    \,,
    \nonumber\\
    &\CB_i \CB_{i+1} = {\bf 1} + q^{-\frac{1}{2}} \CC \, \CA_{i+1} + q^{-1} \CA_{i+1} \CA_{i+1} 
    \,,
    \nonumber\\
    &\CA_i \CB_{i+1} =\CC + q^{-\frac{1}{2}} \CA_{i+1} + q^{\frac{1}{2}} \CA_{i+2}
    \,,
    \label{eq: A3 fusion}
\end{align}
with the subscripts periodic mod 3.

\medskip
\noindent {\bf Monodromy action.}
The monodromy operator can be simplified as
\begin{align}
    \Phi &= (1)(3)(2)(-1)(-3)(-2) 
    \nonumber\\
    &= (2) e^1 e^3 e^2 (-1)(-3)
    \nonumber\\
    &= (x) e^{p+m} e^{p-m} e^x (-p-m)(-p+m)
    \,,
\end{align}
and we check the monodromy action on the line defects --- see appendix \ref{app: A3 monodromy action} for the computation
\begin{align}
    \Phi F(\CA_i) = F(\CA_{i+1}) \Phi
    \;,\;\;
    \Phi F(\CB_i) = F(\CB_{i+1}) \Phi
    \,,
    \label{eq: A3 monodromy action}
\end{align}
which shows the degeneracy independence upon taking the trace
\begin{align}
    \ang{\CA_i}_b = \ang{\CA}_b
    \;,\;\;
    \ang{\CB_i}_b = \ang{\CB}_b
    \,.
    \label{eq: A3 deg indep}
\end{align}

\medskip
\noindent {\bf Map to the Wilson loops.}
Based on~\eqref{eq: A3 deg indep} we can ignore the degeneracy label.  We evaluate $\ang{\CA}_b$ and $\ang{\CB}_b$ from $\CA_1$ and $\CB_1$ respectively as:
\begin{align}
    \ang{\CA}_b
    &=
    i^{\frac{1}{2}} e_b^8
    \int d\s
    \big( 
    e^{2\pi b \s} + e^{-\pi i b^2} e^{-2\pi b m} e^{4\pi b \s}
    \big)
    e^{\pi i (3\s^2 - 2 m \s + 3 m^2)}
    \nonumber\\
    &=\ang{W_{(1)}}_b + q^{\frac{1}{2}} \xi^{-1}\ang{W_{(2)}}_b
    \nonumber\\
    \ang{\CB}_b
    &=
    i^{\frac{1}{2}} e_b^8
    \int d\s
    \big( 
    e^{2\pi b m} e^{-2\pi b \s}
    +
    e^{-\pi i b^2} e^{-4\pi b \s}
    +
    e^{-2\pi i b^2} e^{2\pi b m} e^{-6\pi b \s}
    \big)
    e^{\pi i (3\s^2 - 2 m \s + 3 m^2)}
    \,,
    \nonumber\\
    &=
    \xi \ang{W_{(-1)}}_b
    +
    q^{\frac{1}{2}} \ang{W_{(-2)}}_b
    +
    q \xi \ang{W_{(-3)}}_b
    \,,
    \label{eq: A3 integral}
\end{align}
where we set $x_{\g_m} = m $ the real mass parameter for the flavor symmetry. The common gaussian factor in the integrand implies a 3d $\CN=2$ pure CS theory of $U(1)\times U(1)_f$ gauge group with the level matrix
\begin{align}
    K = 
    \left(
    \begin{array}{c|c}
         3 & -1  \\
         \hline
         -1 & 3
    \end{array}
    \right)
    \,,
\end{align}
where we divided the level matrix by the dynamic and background $U(1)$ factors. This coincides with the gauge theory obtained in~\cite{Go:2025ixu} up to rescaling the mass parameter $m$ of the $U(1)_f$ flavor symmetry.

\medskip
\noindent {\bf The line OPE from Bethe root method.}
By plugging in $\s=-iu$ and $m = -i \n$ together with setting $z=e^{2\pi i u}$ and $y = e^{2\pi i \n}$, the Bethe equation at $b=1$ gives rise to
\begin{align}
    z^3 y^{-1} = -1
    \,.
    \label{eq: A3 wilson loop fusion}
\end{align}
Hence, we find the round three-sphere partition functions by taking $b=1$ from \eqref{eq: A3 integral} as
\begin{align}
    \ang{\CA}
    &= 
    \ang{z^{-1}} - y \ang{z^{-2}}
    =
    \ang{z} - y^{-1} \ang{z^2}\,,
    \nonumber\\
    \ang{\CB} 
    &= 
    y^{-1}\ang{z} - \ang{z^2} + y^{-1} \ang{z^3}
    =
    -\ang{1} + y^{-1} \ang{z} - \ang{z^2}\,.
\end{align}
where we used the Bethe equations~\eqref{eq: A3 wilson loop fusion}. Then, we find a similar homomorphism as in~\cite{Cordova:2016uwk,Neitzke:2017cxz}
\begin{align}
    \ang{\CA \CA} &= \ang{(z-y^{-1}z^2)^2}
    =
    \ang{1} - \ang{-1+y^{-1}z - z^2}
    =
    \ang{{\bf 1}} - \ang{\CB}
    =
    \ang{\CA}\times \ang{\CA}
    \,,
    \nonumber\\
    \ang{\CB  \CB} 
    &= \ang{(-1+y^{-1}z - z^2)^2}
    =
    \ang{1} - (y+y^{-1})\ang{z-y^{-1}z^2} + \ang{2-y^{-1}z + z^2}
    \nonumber\\
    &=
    \ang{{\bf 1}} - \CC \ang{\CA} + \ang{\CA \times \CA}
    =
    \ang{\CB}\times \ang{\CB}
    \,,
    \nonumber\\
    \ang{\CA \CB} &= \ang{\CB \CA}
    =
    \ang{(z-y^{-1}z^2)(-1+y^{-1}z - z^2)}
    =
    (y+y^{-1})\ang{1} -2\ang{z - y^{-1} z^2}
    \nonumber\\
    &=
    \CC\ang{{\bf 1}} - 2\ang{\CA}
    =
    \ang{\CA}\times \ang{\CB}
    \,.
    \label{eq: A3 homo}
\end{align}

\medskip
\noindent {\bf Higher-power monodromy}.
Since $\Phi^3 \sim {\bf 1}$, let us consider the trace of quadratic power of the monodromy which also erases the degeneracy dependence
\begin{align}
    \ang{\CA_i}_b^{(2)} = \ang{\CA}_b^{(2)}
    \;,\;\;
    \ang{\CB_i}_b^{(2)} = \ang{\CB}_b^{(2)}
    \,,
\end{align}
and the map to Wilson loops reads
\begin{align}
    \ang{\CA}_b^{(2)}
    &=
    -i^{\frac{3}{2}} e_b^{16} \int d\s 
    \big(
    e^{2\pi b \s}
    +
    e^{-2\pi b \s}
    \big)
    e^{\pi i (-3\s^2 + 2m\s + 5m^2)}
    =
    \ang{W_{(1)}}_b^{(2)}
    +
    \ang{W_{(-1)}}_b^{(2)}
    \,,
    \nonumber\\
    \ang{\CB}_b^{(2)}
    &=
    -i^{\frac{3}{2}} e_b^{16} \int d\s 
    \big(
    e^{-2\pi b m}
    e^{2\pi b \s}
    +
    e^{\pi i b^2}
    e^{4 \pi b \s}
    +
    e^{2\pi i b^2}
    e^{-2\pi b m}
    e^{6 \pi b \s}
    \big)
    e^{\pi i (-3\s^2 + 2m\s + 5m^2)}
    \nonumber\\
    &=
    y^{-1}
    \ang{ W_{(1)} }_b^{(2)}
    +
    q^{-\frac{1}{2}}
    \ang{ W_{(2)} }_b^{(2)}
    +
    q^{-1}y^{-1}
    \ang{ W_{(3)} }_b^{(2)}
    \,.
\end{align}
At $b=1$, this can be written in terms of the Bethe equation variable as
\begin{align}
    \ang{\CA}^{(2)}
    =
    \ang{z}^{(2)}
    +
    \ang{z^{-1}}^{(2)}
    \;,\;\;
    \ang{\CB}^{(2)}
    =
    y^{-1}\ang{z}^{(2)}
    +
    y\ang{z^{-1}}^{(2)}
    -
    \ang{1}^{(2)}
    \,,
\end{align}
with the equation $z^{-3}y = -1$. Again, this implies the same fusion-preserving homomorphism given in \eqref{eq: A3 homo}:
\begin{align}
    &\ang{\CA  \CA}^{(2)} = \ang{\CA}^{(2)} \times \ang{\CA}^{(2)}
    \,,
    \nonumber\\
    &\ang{\CB  \CB}^{(2)} = \ang{\CB}^{(2)} \times \ang{\CB}^{(2)}
    \,,
    \nonumber\\
    &\ang{\CA  \CB}^{(2)} = \ang{\CA}^{(2)} \times \ang{\CB}^{(2)}
    \,.
\end{align}

\medskip
\noindent {\bf Toward the modular data.}
Let us refer back to the first power monodromy trace. Since the gauge theory is an abelian CS theory at level 3, a natural set of Wilson loops that describe the anyons are
\begin{align}
    W_{(\a)}(z) = z^{-\a}
    \;,\;\;
    \text{for}\;
    \a=0,1,2\,.
    \label{eq: A3 wilson loops}
\end{align}
Then, by turning off the flavor fugacity $y=1$, we find three Bethe vacua from \eqref{eq: A3 wilson loop fusion} as
\begin{align}
    z^{(\b)} = - e^{\frac{2\pi i}{3}\b}
    \;,\;\;
    \text{for}\;
    \b=0,1,2\,.
\end{align}
Equipped with the handle gluing and fibering operators from the 3d A-model method
\begin{align}
    \CH(z) = 3
    \;\;,\;\;\;
    \CF(z) = e^{-\frac{3}{4\pi i}(\log(z))^2}
    \,,
\end{align}
together with the formula for computing $S$- and $T$-matrix (normalized as $T_{00}=1$)
\begin{align}
    &
    S_{\a\b}
    =
    W_{(\a)}(z^{(\b)}) S_{0\b}
    \;\;,\;\;\;
    T_{\a\b} = \d_{\a\b} \CF(z^{(0)})/\CF(z^{(\a)})
    \,,
\end{align}
we get the modular matrices 
\begin{align}
    S = -\frac{1}{\sqrt{3}}
    \begin{pmatrix}
        -1 & 1 & -1 \\
        1 & e^{2\pi i (1/6)} & e^{2\pi i (1/3)} \\
        -1 & e^{2\pi i (1/3)} & e^{2\pi i (1/6)}
    \end{pmatrix}
    \;,\;\;
    T = 
    \begin{pmatrix}
        1 & 0 & 0 \\
        0 & e^{2\pi i (2/3)} & 0 \\
        0 & 0 & e^{2\pi i (2/3)}
    \end{pmatrix}
    \,,
    \label{eq: naive A3}
\end{align}
which are compatible with that of the affine $su(2)$ VOA at admissible level $-4/3$ with central charge $c=-6$ up to an overall sign of $S$ and a phase factor of $T$:
\begin{align}
    S = \frac{1}{\sqrt{3}}
    \begin{pmatrix}
        -1 & 1 & -1 \\
        1 & e^{2\pi i (1/6)} & e^{2\pi i (1/3)} \\
        -1 & e^{2\pi i (1/3)} & e^{2\pi i (1/6)}
    \end{pmatrix}
    \;,\;\;
    T = 
    e^{\frac{2\pi i}{24} (-6)}
    \begin{pmatrix}
        1 & 0 & 0 \\
        0 & e^{2\pi i (2/3)} & 0 \\
        0 & 0 & e^{2\pi i (2/3)}
    \end{pmatrix}
    \,,
\end{align}
This example illustrates a systematic way of computing the full S-matrix, whereas the usual approach computes only the first row. This is because the Wilson loops \eqref{eq: A3 wilson loops} flow to the simple objects in the TQFT. We further develop this discussion in the next section.

\section{Simple lines and modular data} \label{sec: 3}
In this section, we find UV line operators of 3d $\CN=2$ gauge theories arising from the $U(1)_r$ twisted circle reduction of the $G$ theories which are mapped to simple objects~\cite{Moore:1988qv,Turaev:1994xb}, or anyons, of the resulting IR topological phases. These are referred to as {\it simple lines} in the literature~\cite{Creutzig:2024ljv,Gang:2024loa,Kim:2024dxu,Creutzig:2026ajk}.  Based on methods developed in a series of earlier works~\cite{Cho:2020ljj,Gang:2021hrd,Gang:2024loa,Jeong:2025xid}, we propose an algorithm for constructing the modular $S$- and $T$-matrices of the TQFT by imposing non-trivial necessary conditions on simple lines via supersymmetric indices.

\begin{table}
    \centering
    \begin{tabular}{c|ccccccc}
       $G$ & $A_{2m}$ & $A_{2m+1}$ &  $D_{2m}$ &  $D_{2m+1}$ &  $E_{6}$ &  $E_{7}$ &  $E_{8}$
        \\
        \hline
        $p$ & $2m+3$ & $m+2$ & $m$ & $2m+1$ & $7$ & $5$ & $8$
    \end{tabular}
    \caption{
    Monodromy periodicity $\Phi^p \sim {\bf 1}$ of the $G$ theories.
    }\label{tab: period}
\end{table}

\medskip\noindent{\bf Higher-power monodromy}.
As an application of the algorithm, we examine the trace with higher powers of the monodromy operator $\Phi$; see~\cite{Cecotti:2010fi,Iqbal:2012xm,Cecotti:2015lab} for previous discussions, and~\cite{Kim:2024dxu} for recent work computing families of vacuum characters for several $G$ theories. In our case, the trace $\Tr \Phi^n$ provides a 3d $\CN=2$ ACSM theory arising from a multi-wrapping $U(1)_r$ twisted circle reduction of the 4d $G$ theory ~\cite{Go:2025ixu}.  The 3d ACSM theory is associated to IR TQFT(s) which we denote by
\begin{align*}
    \text{TFT}[G^{(n),\n}].
\end{align*}
If $\n=-$ or $\n=+$, the TQFT is non-unitary and understood to arise from the A- or B-twists of an IR rank-0 SCFT respectively.  If $\n$ is absent the TQFT is unitary and arises in the IR directly from RG flow.

Because $\Phi^p \sim {\bf 1}$ with $p$ in Table \ref{tab: period}, the powers of $\Phi$ are periodic with period $p$. Thus ${\rm TFT}[G^{(n),\n}]$ forms a finite orbit as one varies $n$; only becoming semi-simple when ${\rm gcd}(n,p)=1$. It is believed that the modular data in this orbit are related by Galois conjugation. Compared to previous methods~\cite{Go:2025ixu}, which only yielded partial modular data, we determine the full modular data for various examples and confirm that the modular data of ${\rm TFT}[G^{(n),\n}]$ form a Galois orbit for each fixed $G$ whose BPS quiver has at most six nodes.

\subsection{Algorithm for determining the UV lines}
Suppose we have a 3d $\CN=2$ gauge theory that admits a TQFT phase in the IR. Since the gauge theories in our cases are ACSM theories, natural candidates for the simple lines in the TQFT phase are supersymmetric Wilson lines $W_Q$ of charge $Q=(Q_1,\cdots,Q_r)$ under the $U(1)^r$ gauge group. In the 3d A-model construction on a supersymmetric backgound
\begin{align}
    S^1 \;\;\overset{p}{\to}\;\; \CM_{g,p} \;\;\to\;\; \S_g\;,
\end{align}
the Wilson line wraps the $S^1$ fiber, becoming a loop operator. The partition function with its insertion can be computed as~\cite{Closset:2017zgf}
\begin{align}
    \ang{W_Q}_{\CM_{g,p}}
    =
    \sum_{\a}
    W_Q(z^{(\a)}) (\CH_\a)^{g-1} (\CF_\a)^{p}
    \,,
\end{align}
where the sum is over the Bethe vacua $z^{(\a)}$ labeled by $\a$. The components in this equation are given by~\footnote{As explained in appendix \ref{app: A-model}, the handle gluing and fibering operators here should be understood as the shifted ones $(\tilde{\CH},\tilde{\CF})$ by some possible R-symmetry mixing due to superpotential terms. We denote them without tilde through the subsequent discussion for notational simplicity.} 
\begin{align}
    &W_Q(z) := \prod_{i=1}^r z_i^{-Q_i}
    \;,\;\;
    \CH_\a := \CH(z^{(\a)}) 
    \;,\;\;
    \CF_\a := \CF(z^{(\a)})
    \,.
    \label{eq: Bethe components}
\end{align}

Let us denote the simple line in the IR TQFT by $\mathfrak{L}_\a$, and its corresponding Wilson line charge by $Q^{(\a)}$. Then, the $S^3\cong\CM_{0,1}$ expectation value of $\mathfrak{L}_\a$ computes $S_{0\a} = \frac{\z_\a}{\sqrt{\CH_\a}}$ up to a phase factor $\z_\a$. This suggests a necessary condition on $Q^{(\a)}$:
\begin{align}
    |\ang{W_{Q^{(\a)}}}| = \frac{1}{\sqrt{|\CH_\a}|}
    \,.
    \label{eq: Q cond}
\end{align}
Going forward we will suppress the subscript $\CM_{g=0,p=1}$ on the $S^3$ expectation value. Note that if we map a trivial Wilson line to the trivial simple line $\mathfrak{L}_0 = {\bf 1}$, then we have $|\ang{W_{Q={\bf 0}}}| = |Z_{S^3}|= 1/\sqrt{|\CH_0|} $ which distinguishes the true vacuum $z^{(\a=0)}$ whenever $\CH_0$ differs from the other $\CH_\a$. 

In our approach we first collect a set of candidate charges, $\widetilde{\mathfrak{Q}}^{(\a)}$, for each $\a$ by scanning $Q^{(\a)}$:
\begin{align}
    \widetilde{\mathfrak{Q}}^{(\a)} := \{ Q^{(\a)}\ |\ Q^{(\a)}\ \rm{satisfies }\ \eqref{eq: Q cond}\}
    \,.
\end{align}
$\widetilde{\mathfrak{Q}}^{(\a)}$ is then further reduced by requiring the superconformal index to satisfy the following simple line conditions (see appendix \ref{app: A-model})~\cite{Gang:2024loa} :
\begin{align}
    &\ang{W_{Q^{(\a)}}^\pm}_{\rm sci} = 0, \; (\text{or}\; \pm q^{\mathbb{Z}/2})
    \,,
    \nonumber\\
    &\ang{W_{Q^{(\a)}}^+ W_{Q^{(\a)}}^-}_{\rm sci} = 1
    \ .
    \label{eq: SCI cond}
\end{align}
In the superconformal index background the Wilson line that wraps the $S^1$ fiber in $S^2 \times S^1$ carries an additional superscript $\pm$, labeling the two supersymmetric insertion points at the north and south poles  respectively~\cite{Kapustin:2009kz,Agarwal:2016pjo}. The first condition in \eqref{eq: SCI cond} implies that only the trivial line insertion results in a non-vanishing index.  The second condition measures the overlap between $\mathfrak{L}_\a \times \overline{\mathfrak{L}}_\a = {\bf 1} + \cdots$ and the trivial line ${\bf 1}$, thus, becoming 1.  Therefore we are left with the candidate charges:
\begin{align}
    \mathfrak{Q}^{(\a)} := \{ Q^{(\a)}\ |\ Q^{(\a)}\ \rm{satisfies }\ \eqref{eq: Q cond}\ \rm{and}\ \eqref{eq: SCI cond}\}
    \,.
\end{align}
There is a one-to-one correspondence between the simple object $\mathfrak{L}_\a$ and the Bethe vacuum $z^{(\a)}$~\cite{Gang:2021hrd}. If the values of $\CH_\a$ are distinct, this simple object-Bethe root map is determined via the condition \eqref{eq: Q cond}. Then, we pick a representative charge $Q^{(\a)}$ from each $\mathfrak{Q}^{(\a)}$ and compute the $S$-matrix with the formula~\cite{Cho:2020ljj}
\begin{align}
    S_{\a\b} = \e_\a W_{Q^{(\a)}}(z^{(\b)}) S_{0\b}
    \;,\;\;
    (S_{0\b})^{-2} = \CH_\b
    \,,
    \label{eq: S from W}
\end{align}
where $\e_\a = \pm 1$ is a possible sign ambiguity that we should consistently pick. Note that, with $\b=0$, the signs of $S_{\a0}$ is fixed once $\e_\a$ is given and by requiring $S_{00}>0$. Enforcing symmetry of $S_{\a\b}$ and non-negativity of fusion coefficients $N_{\a\b}^\g$ obtained via the Verlinde formula
\begin{align}
    N_{\a\b}^\g
    =
    \sum_{\s}
    \frac{S_{\a\s} S_{\b\s} S^*_{\s\g} }{S_{0\s}}
\end{align}
completely fixes $\e_\a$ in all the examples. The result is that we determine the $S$-matrix from several non-trivial necessary conditions along with a conjectural map from UV Wilson lines to IR simple lines:
\begin{align}
    \e_\a W_{Q^{(\a)}}(z) \; \to \; \mathfrak{L}_\a
    \,.
    \label{eq: W to L}
\end{align}
We also compute the $T$-matrix, normalized by $T_{00}=1$, from the fibering operator as
\begin{align}
    T_{\a\b} = \d_{\a\b} \CF_0 / \CF_\a
    \,,
    \label{eq: T from F}
\end{align}
and determine the central charge $c$ mod 8 from the $SL(2,\mathbb{Z})$ group relation applied to projective representations: $(ST)^3 = e^{\frac{\pi i}{4} c} S^2$. If we had instead required $S_{00}<0$, the effect would be to take $S \to -S$, shifting $c$ by $4\,\text{mod}\,8$.

We make two remarks concerning our approach. First, the procedure requires the simple object-Bethe root map. This is uniquely determined if $\CH_\a \neq \CH_\b$ for $\a \neq \b$. However, if $\CH_\a = \CH_\b$ for some $\a \neq \b$, it is not uniquely determined and there may arise multiple choices of seemingly valid modular data. In such cases, we further use a consistency condition
\begin{align}
    (S^{-1})_{\a\b} = e^{2\pi i \d} \CF_\a \CF_\b \ang{\mathfrak{L}_\a\,\mathfrak{L}_\b}
    \,,
\end{align}
to find valid maps between simple objects and Bethe vacua, presenting all possible consistent results in the remaining examples. Here a possible overall phase $\d$ from framing is fixed such that $S_{00}>0$. Second, the simple line may not be mapped to a supersymmetric Wilson line consisting of a monomial term as in \eqref{eq: Bethe components}.  Monomial  descriptions for the UV lines were observed in~\cite{Gang:2023rei, Gang:2024loa} but are not required. We identify multinomial descriptions for the candidate UV line whenever a monomial which computes a valid $S$-matrix cannot be found. 

\medskip\noindent {\bf Summary of the algorithm}.
For clarity, we summarize our algorithm for determining the UV lines that flow to the simple lines as well as the modular data of the IR TQFT phase:
\begin{enumerate}
    \item Find candidate Wilson loop charges $\widetilde{\mathfrak{Q}}^{(\a)} = \{  Q^{(\a)} \}$ for each $\a$ from the necessary condition: $|\ang{W_{Q^{(\a)}}}| = \frac{1}{\sqrt{|\CH_\a|}}$.

    \item Reduce $\widetilde{\mathfrak{Q}}^{(\a)}\to\mathfrak{Q}^{(\a)}$ by selecting $Q^{(\a)}$ that satisfy the simple line condition in the superconformal index background: $\ang{ W^+_{Q^{(\a)}} W^-_{Q^{(\b)}} }_{\rm sci} = \d_{\a\b}$.

    \item Among the remaining charges in each $\mathfrak{Q}^{(\a)}$, pick $Q^{(\a)}$ and compute the $S$-matrix $(S_{00}>0)$ via \eqref{eq: S from W}, with a sign choice $\e_\a$ attached to each Wilson loop $W_{Q^{(\a)}}$.

    \item Determine $\e_\a$ by imposing $S_{\a\b} = S_{\b\a}$ and non-negative fusion coefficients $N_{\a\b}^\g \in \mathbb{Z}_{\geq 0}$.

    \item Compute the $T$-matrix (normalized as $T_{00}=1$) from the fibering operators \eqref{eq: T from F}, and determine the central charge $c\, \text{mod}\, 8$ from $(ST)^3 = e^{\frac{\pi i}{4}c} S^2$.

\end{enumerate}

In the following examples, we consider 3d $\CN=2$ abelian gauge theories arising from the $U(1)_r$ twisted circle reduction of the $G$ theories, which flow to the 3d TQFTs in the IR. We determine the UV Wilson lines with signs that flow to the simple lines as described in \eqref{eq: W to L} and present them as an ordered set
\begin{align}
    \L = \big\{ \e_0 W_{Q^{(0)}}(z)\;,\; \e_1 W_{Q^{(1)}}(z) , \cdots \big\}
\end{align}
accordingly with the order of the simple objects in the computed modular $S$ and $T$ matrices.

\subsection{Examples}
We apply our algorithm to several $\text{TFT}[G^{(n),\n}]$ theories to determine their simple lines $\L$ and modular data $\{S,h,c\}$ where $h = (h_0,h_1,\cdots)$ denotes the topological spins defined modulo 1 via $T_{\a\a}=e^{2\pi i h_\a}$, for cases in which the BPS quiver of the $G$ theory has at most six nodes.

The 3d $\CN=2$ UV gauge theory for $\text{TFT}[G^{(n),\n}]$ obtained from $\Tr \Phi^n$ is characterized by CS level matrix $K=(K_{ij})$, charge matrix $C=(C_{iJ})$ of the chiral multiplets $\{\Phi_J\}$, and a set of dressed half-BPS monopole operators $\CV=\big\{ \prod_{J}\phi_J^{d_{J}} V_{\mathfrak{m}} \big\}$ as superpotential terms; $\phi_J$ is the scalar component of $\Phi_J$ with some dressing number $d_J \in \mathbb{Z}_{\geq 0}$, and $\mathfrak{m}$ is the magnetic flux of the bare monopole operator $V_{\mathfrak{m}}$. $K$ and $C$ are read off from the integral, while a compatible set of monopole superpotential terms $\CV$ is inferred from $K$ and $C$ by requiring half-BPS conditions~\cite{Gaiotto:2024ioj,Go:2025ixu,Kim:2025rog,Nishinaka:2025ytu}
\begin{align}
    \Tr \Phi^n = Z_{S_b^3}^{{\rm TFT}[G^{(n),\n}]}
    \;\;
    \to
    \;\;
    \{K,C,\CV\}
    \,.
\end{align}

In the following, we only consider $n \leq \lfloor p/2\rfloor$, since $\text{TFT}[G^{(n),\n}]$ and $\text{TFT}[G^{(p-n),\n}]$ are related by orientation reversal modulo gravitational, and possibly background CS, couplings~\cite{Go:2025ixu}. The modular data for orientation reversed theories are related by complex conjugation and taking $c\to-c\mod 8$.

The trace of the first power monodromy computes the partition function of a 3d TQFT mediating the SCFT/VOA correspondence~\cite{Go:2025ixu}. Hence, the extracted modular data is expected to be compatible with that of the VOA. However, since our algorithm imposes non-negativity on the fusion coefficients, we will get a different modular S-matrix from the SCFT/VOA result whenever the latter contains negative fusion coefficients.  In this setting ``compatible'' means the two S-matrices will be related by a signed permutation on the bases.

For all cases except the fermionic one, we check that the modular data we find is valid by referencing the known classification up to rank 12 by Ng-Rowell-Wen (NRW) \cite{ng2025classificationmodulardatarank}. In the classification each modular data is given a label: 
$$
r^{\rm{Ord}(T),\rm{fp}}_{c,D}
$$
which we will refer to as \textit{modular labels} which we will list in each theory we study. Here $r$ is the rank, $\rm{Ord}(T)$ is the order of the $T$-matrix, $c$ is the central charge defined mod $8$, $D$ is the total quantum dimension,
\begin{equation}
D=\sqrt{\sum_{\a} ( S_{0\a} / S_{00} )^2},
\end{equation}
 and $\rm{fp}$ is the ``quantum fingerprint'' defined to be the first 3 digits of
\begin{equation}
\left|\sum_{\a}\left(h_{\a}^2-\frac14\right)\frac{S_{0\a}}{S_{00}}\right|.
\end{equation}
We summarize the results in Table \ref{tab: MTC}.
\begin{table}[tbp]
\renewcommand{\arraystretch}{1.1}
    \centering
    \begin{tabular}{|c|c|c|}
       \hline
       $G^{(n),\n}$ & \text{modular label} & \text{comment} \\
       \hline
       \hline
       $A_2^{(1),\pm}$ & $2_{\frac{2}{5},1.381}^{5,120}/ 2_{\frac{38}{5},1.381}^{5,491}  $ & \scriptsize $osp(1|2)_1 / \texttt{-} M(2,5)$ \\
       \hline
       $A_2^{(2)}$ & $2_{\frac{26}{5},3.618}^{5,720}$ & \scriptsize Fibonacci* \\
       \hline
       $A_3^{(1)}$ & $3_{6,3}^{3,138}$ & \\
       \hline
       $A_4^{(1),\pm}$ & $3_{\frac{4}{7},1.841}^{7,953}/3_{\frac{44}{7},2.862}^{7,531}$ & \scriptsize $osp(1|2)_2 / M(2,7)$ \\
       \hline
       $A_4^{(2),\pm}$ & $3_{\frac{44}{7},2.862}^{7,531}/3_{\frac{4}{7},1.841}^{7,953}$ & \scriptsize $ M(2,7) / osp(1|2)_2$ \\
       \hline
       $A_4^{(3)}$ & $3_{\frac{48}{7},9.295}^{7,790}$ & \scriptsize $ (A_1,5)_{\frac{1}{2}}$* \\
       \hline
       $A_5^{(1)}$ & \rm{fermionic} & \rm{see} \eqref{eq: A5 ST} \\
       \hline
       $A_6^{(1),\pm}$ & $4_{\frac{2}{3},2.319}^{9,199}/4_{\frac{14}{3},5.445}^{9,544}$ & \scriptsize $osp(1|2)_3 / \texttt{-} M(2,9)$ \\
       \hline
       $A_6^{(2),\pm}$ & $4_{\frac{10}{3},5.445}^{9,616}/4_{\frac{22}{3},2.319}^{9,549}$ & \\
       \hline
    \end{tabular}
    \;
    \begin{tabular}{|c|c|c|}
       \hline
       $G^{(n),\n}$ & \text{modular label} & \text{comment} \\
       \hline
       \hline
       $A_6^{(4)}$ & $4_{\frac{14}{3},19.23}^{9,614}$ & \\
       \hline
       $D_4^{(1)}$ & $4_{0,4}^{2,750},4_{0,4}^{2,250},4_{4,4}^{2,250}$ & \\
       \hline
       $D_5^{(1)}$ & $5_{4,5}^{5,210}$ & \\
       \hline
       $D_5^{(2)}$ & $5_{0,5}^{5,110}$ & \\
       \hline
       $D_6^{(1)}$ & $9_{0,9}^{3,113}$ & \\
       \hline
       $E_6^{(1),\pm}$ & $5_{\frac{6}{7},2.155}^{7,342}/5_{\frac{26}{7},4.501}^{7,408}$ & \scriptsize $/ \texttt{-}W_3(3,7)$ \\
       \hline
       $E_6^{(2)}$ & $5_{\frac{38}{7},35.34}^{7,386}$ & \\
       \hline
       $E_6^{(3),\pm}$ & 
       $5_{\frac{30}{7},4.501}^{7,125}/5_{\frac{50}{7},2.155}^{7,255}$ & \\
       \hline
    \end{tabular}
    \caption{
    The modular labels of the 3d TQFTs ${\rm TFT}[G^{(n),\n}]$ arising from the multi-power monodromy traces $\Tr \Phi^n$. We comment on the alternative VOA identification, where the minus sign and $*$ denote an overall sign flip $S\to-S$ and complex conjugate respectively. The modular data for every $G$ are included in a Galois orbit. We could not identify the modular data of the $A_5$ theory, since the resulting TQFT is fermionic.
    }\label{tab: MTC}
\end{table}

\subsubsection{\texorpdfstring{$A_2$}{A2} theory}
The monodromy of the $A_2$ theory has periodicity $p=5$, thus we only consider $n=1$ and $2$.
\paragraph{\texorpdfstring{$\boldsymbol{{\rm TFT}[A_2^{(1),\n}]}$}{A2(1)}}
As we discussed in the previous section, once-wrapping $U(1)_r$ twisted circle reduction of $A_2$ theory gives rise to the GY rank-0 SCFT:
\begin{align}
    K=(2)
    \;,\;
    C=(1)
    \;,\;
    \CV=\emptyset
    \,,
\end{align}
where the topological symmetry $U(1)_{T_1}$ is mapped to the axial symmetry $U(1)_A$ of the $\CN=4$ supersymmetry, i.e., $A = T_1$ in terms of the respective generators.

We identify two Wilson lines that flow to the two simple lines in ${\rm TFT}[A_2^{(1),\n}]$, computing the modular data to be
\begin{align}
    &\n=-
    \;:\;
    \L = \{ 1 , z_1^{-1} \}
    \nonumber\\
    &\qquad\qquad\to\quad
    S=
    \frac{2}{\sqrt{5}}
    \begin{pmatrix}
        \sin \frac{2\pi}{5} & -\sin \frac{\pi}{5} \\
        -\sin \frac{\pi}{5} & -\sin \frac{2\pi}{5}
    \end{pmatrix}
    ,\;
    h=\Big(0,\frac{4}{5}\Big)
    ,\;
    c=\frac{38}{5}
    \,.
    \label{eq: A2(1)A}
    \\
    &\n=+
    \;:\;
    \L = \{ 1 , -z_1 \}
    \nonumber\\
    &\qquad\qquad\to\quad
    S=
    \frac{2}{\sqrt{5}}
    \begin{pmatrix}
        \sin \frac{2\pi}{5} & -\sin \frac{\pi}{5} \\
        -\sin \frac{\pi}{5} & -\sin \frac{2\pi}{5}
    \end{pmatrix}
    ,\;
    h=\Big(0,\frac{1}{5}\Big)
    ,\;
    c=\frac{2}{5}
    \,.
    \label{eq: A2(1)B}
\end{align}
The modular labels are $2^{5,491}_{\frac{38}{5},1.381}$ and $2^{5,120}_{\frac{2}{5},1.381}$ respectively. Note that \eqref{eq: A2(1)A} is the modular data of affine $E_{7\frac{1}{2}}$ at level 1, and could also correspond to that of the $M(2,5)$ Virasoro minimal model with $c=-22/5$ upon an overall sign flip of $S$. The modular data in \eqref{eq: A2(1)B} is affine $osp(1|2)$ at level 1.

\paragraph{\texorpdfstring{$\boldsymbol{{\rm TFT}[A_2^{(2)}]}$}{A2(2)}}
The quadratic monodromy trace gives rise to a 3d $\CN=2$ ACSM theory
\begin{align}
    K=
    \begin{pmatrix}
        0 & 2 \\
        2 & 0
    \end{pmatrix}
    ,\;
    C=I_{2}
    ,\;
    \CV=\{
    \phi_1^2 V_{(0,-1)}
    ,
    \phi_2^2 V_{(-1,0)}
    \}
    \,.
\end{align}
The superpotential terms $\CV$ completely break the topological symmetries and the theory flows to the unitary gapped theory: ${\rm TFT}[A_2^{(2)}]$~\cite{Go:2025ixu}. We identify the two Wilson lines mapped to the simple lines and compute the modular data to be
\begin{align}
    \L = \{ 1 , -z_1 \}
    \;\;\to\;\;
    S=
    \frac{2}{\sqrt{5}}
    \begin{pmatrix}
        \sin \frac{\pi}{5} & \sin \frac{2\pi}{5} \\
        \sin \frac{2\pi}{5} & -\sin \frac{\pi}{5}
    \end{pmatrix}
    ,\;
    h=\Big(0,\frac{3}{5}\Big)
    ,\;
    c=\frac{26}{5}
    \,,
\end{align}
which are those of affine $F_4$ at level 1 or, equivalently, the modular representation of the conjugate Fibonacci MTC. The modular label is $2^{5,720}_{\frac{26}{5},3.618}$.

\subsubsection{\texorpdfstring{$A_3$}{A3} theory}
Since $\Phi^3 \sim {\bf 1}$ for the $A_3$ theory, we only consider the first power monodromy trace.
\paragraph{\texorpdfstring{$\boldsymbol{{\rm TFT}[A_3^{(1)}]}$}{A3(1)}}
As we obtained in the previous section, the 3d gauge theory becomes a pure $U(1) \times U(1)_f $ CS theory where $U(1)_f$ corresponds to the $SU(2)$ flavor symmetry of the $A_3$ theory
\begin{align}
    K
    =
    \left(
    \begin{array}{c|c}
        3 & -1 \\
        \hline
        -1 & 3
    \end{array}
    \right)
    \,,
\end{align}
which flows to a unitary TQFT ${\rm TFT}[A_3^{(1)}]$ in the IR. We detemine the simple lines and modular data to be
\begin{align}
    \L = \{ 1 , -z_1, z_1^2 \}
    \;\;\to\;\;
    S=
    \frac{1}{\sqrt{3}}
    \begin{pmatrix}
        1 & 1 & 1 \\
        1 & e^{\frac{4\pi i}{3}} & e^{\frac{2\pi i}{3}} \\
        1 & e^{\frac{2\pi i}{3}} & e^{\frac{4\pi i}{3}}
    \end{pmatrix}
    ,\;
    h=\Big(0,\frac{2}{3},\frac{2}{3}\Big)
    ,\;
    c=6
    \,.
    \label{eq: A3 STc}
\end{align}
The modular label is $3^{3,138}_{6,3}$. Although \eqref{eq: naive A3} and \eqref{eq: A3 STc} are related by sign rephasings of the basis and therefore define equivalent modular representations, only the latter yields non-negative integral coefficients under the Verlinde formula. We therefore use \eqref{eq: A3 STc} as the modular $S$-matrix in the simple-line basis of the TQFT.

\subsubsection{\texorpdfstring{$A_4$}{A4} theory}
The monodromy operator of $A_4$ theory has periodicity $p=7$, so we only consider $\Phi^n$ for $n=1,2,3$.
\paragraph{\texorpdfstring{$\boldsymbol{{\rm TFT}[A_4^{(1),\n}]}$}{A4(1)}}
The trace $\Tr \Phi$ of $A_4$ theory reads 3d $\CN=2$ ACSM theory as
\begin{align}
    K
    =
    \begin{pmatrix}
        -1 & 2 \\
        2 & 0
    \end{pmatrix}
    ,\;
    C = I_2
    ,\;
    \CV = \{
    \phi_1^2 V_{(0,-1)}
    \}
    \,,
\end{align}
which flows to a rank-0 SCFT with the $\CN=4$ axial symmetry identified as $A=T_1$. Then, the 3d A-/B-twists give rise to non-unitary TQFTs ${\rm TFT}[A_4^{(1),\mp}]$ with simple line and modular data determined by our method to be
\begin{align}
    &\n=-
    \;:\;
    \L = \{ 1 , -z_1^{-1} , -z_1^{-1}z_2 \}
    \nonumber\\
    &\qquad\to\quad
    S=
    \frac{2}{\sqrt{7}}
    \begin{pmatrix}
        \sin \frac{2\pi}{7} & \sin \frac{\pi}{7} & -\sin \frac{3\pi}{7} \\
        \sin \frac{\pi}{7} & \sin \frac{3\pi}{7} & \sin \frac{2\pi}{7} \\
        -\sin \frac{3\pi}{7} & \sin \frac{2\pi}{7} & - \sin \frac{\pi}{7}
    \end{pmatrix}
    ,\;
    h=\Big(0,\frac{4}{7},\frac{5}{7}\Big)
    ,\;
    c=\frac{44}{7}
    \,.
    \label{eq: A4(1)A}
    \\
    &\n=+
    \;:\;
    \L = \{ 1 , -z_1, -z_2 \}
    \nonumber\\
    &\qquad\to\quad
    S=
    \frac{2}{\sqrt{7}}
    \begin{pmatrix}
        \sin \frac{3\pi}{7} & \sin \frac{\pi}{7} & -\sin \frac{2\pi}{7} \\
        \sin \frac{\pi}{7} & \sin \frac{2\pi}{7} & \sin \frac{3\pi}{7} \\
        -\sin \frac{2\pi}{7} & \sin \frac{3\pi}{7} & - \sin \frac{\pi}{7}
    \end{pmatrix}
    ,\;
    h=\Big(0,\frac{3}{7},\frac{1}{7}\Big)
    ,\;
    c=\frac{4}{7}
    \,.
    \label{eq: A4(1)B}
\end{align}
The modular labels are $3^{7,531}_{\frac{44}{7},2.862}$ and $3^{7,953}_{\frac{4}{7},1.841}$ respectively. Up to $c$ mod $8$, \eqref{eq: A4(1)A} is the modular data of $M(2,7)$ minimal model with $c=-68/7$, while \eqref{eq: A4(1)B} is that of the affine $osp(1|2)$ at level 2.

\paragraph{\texorpdfstring{$\boldsymbol{{\rm TFT}[A_4^{(2),\n}]}$}{A4(2)}}
The trace $\Tr \Phi^2$ reads 3d $\CN=2$ ACSM theory as
\begin{align}
    K
    =
    \begin{pmatrix}
        3 & -2 \\
        -2 & 1
    \end{pmatrix}
    ,\;
    C = I_2
    ,\;
    \CV = \{
    \phi_1^2 V_{(0,1)}
    \}
    \,,
\end{align}
which also flows to $\CN=4$ rank-0 SCFT with axial symmetry generated by $A = - T_1$. Then, at the 3d A-/B-twist sectors capture non-unitary TQFTs ${\rm TFT}[A_4^{(2),\n}]$ with simple lines and modular data determined to be
\begin{align}
    &\n=-
    \;:\;
    \L = \{ 1 , -z_1 , z_1^2 z_2^{-1} \}
    \nonumber\\
    &\qquad\to\quad
    S=
    \frac{2}{\sqrt{7}}
    \begin{pmatrix}
        \sin \frac{3\pi}{7} & \sin \frac{\pi}{7} & -\sin \frac{2\pi}{7} \\
        \sin \frac{\pi}{7} & \sin \frac{2\pi}{7} & \sin \frac{3\pi}{7} \\
        -\sin \frac{2\pi}{7} & \sin \frac{3\pi}{7} & - \sin \frac{\pi}{7}
    \end{pmatrix}
    ,\;
    h=\Big(0,\frac{3}{7},\frac{1}{7}\Big)
    ,\;
    c=\frac{4}{7}   
    \,.
    \label{eq: A4(2)A}
    \\
    &\n=+
    \;:\;
    \L = \{ 1 , -z_1^{-1}, z_1 z_2^{-1} \}
    \nonumber\\
    &\qquad\to\quad
    S=
    \frac{2}{\sqrt{7}}
    \begin{pmatrix}
        \sin \frac{2\pi}{7} & \sin \frac{\pi}{7} & -\sin \frac{3\pi}{7} \\
        \sin \frac{\pi}{7} & \sin \frac{3\pi}{7} & \sin \frac{2\pi}{7} \\
        -\sin \frac{3\pi}{7} & \sin \frac{2\pi}{7} & - \sin \frac{\pi}{7}
    \end{pmatrix}
    ,\;
    h=\Big(0,\frac{4}{7},\frac{5}{7}\Big)
    ,\;
    c=\frac{44}{7}
    \,.
    \label{eq: A4(2)B}
\end{align}
The modular labels are  $3^{7,953}_{\frac{4}{7},1.841}$ and $3^{7,531}_{\frac{44}{7},2.862}$ respectively. These modular data are affine $osp(1|2)$ at level 2 and $M(2,7)$ respectively, as previously obtained in \eqref{eq: A4(1)B} and \eqref{eq: A4(1)A}.

\paragraph{\texorpdfstring{$\boldsymbol{{\rm TFT}[A_4^{(3)}]}$}{A4(3)}}
The cubic monodromy trace $\Tr \Phi^3$ gives rise to a 3d $\CN=2$ ACSM theory
\begin{align}
    &\qquad\qquad\qquad
    K = 
    \begin{pmatrix}
        0 & 2 & 0 & 0 \\
        2 & 0 & -2 & 0 \\
        0 & -2 & 0 & 2 \\
        0 & 0 & 2 & 0
    \end{pmatrix}
    \;\;,\;\;\;
    C = I_4
    \;,
    \nonumber\\
    &\CV =
    \{
    \phi_1^2 V_{(0,-1,0,-1)} ,
    \phi_2^2 V_{(-1,0,0,0)} ,
    \phi_3^2 V_{(0,0,0,-1)} ,
    \phi_4^2 V_{(-1,0,-1,0)}
    \}
    \,,
\end{align}
which flows directly to a unitary TQFT, ${\rm TFT}[A_4^{(3)}]$ with simple lines and modular data are determined to be
\begin{align}
    \L = \{1 , z_1 z_3^{-1},-z_2\}
    \;\to\;
    S=
    \frac{2}{\sqrt{7}}
    \begin{pmatrix}
        \sin \frac{\pi}{7} & \sin \frac{2\pi}{7} & \sin \frac{3\pi}{7} \\
        \sin \frac{2\pi}{7} & -\sin \frac{3\pi}{7} & \sin \frac{\pi}{7} \\
        \sin \frac{3\pi}{7} & \sin \frac{\pi}{7} & - \sin \frac{2\pi}{7}
    \end{pmatrix}
    ,\;
    h=\Big(0,\frac{1}{7},\frac{5}{7}\Big)
    ,\;
    c=\frac{48}{7}
    \,.
\end{align}
The modular label is $3^{7,790}_{\frac{48}{7},9.295}$. These are the complex conjugate of $(A_1,5)_{\frac{1}{2}}$ modular data~\cite{Rowell:2007dge}.

\subsubsection{\texorpdfstring{$A_5$}{A5} theory}

\paragraph{\texorpdfstring{$\boldsymbol{{\rm TFT}[A_5^{(1)}]}$}{A5(1)}}
Based on the SCFT/VOA result we expect the TQFT for this case to be fermionic \cite{Beem:2017ooy, adamovic2019classificationirreduciblemodulesbershadskypolyakov, Auger:2019gts}. Thus, we now have $(\CF^{(\a)})^{-2}=T_{\a\a}^2$ which implies our computational results for the topological spins are defined mod $\frac12$, and the property $(ST)^3=e^{\frac{2\pi ic}{8}}S^2$ does not apply since the modular representation is that of a subgroup generated by $S$ and $T^2$. Also, the results in NRW focus on bosonic data so we can no longer use them as a guide. Nevertheless, we still extract modular data in some capacity. There exists a proposed semi-classification of fermionic modular data up to ``unresolved cases'' \cite{PhysRevB.108.115103} but our $(S,T)$ pair does not appear in their findings.

The UV ACSM theory is given by 
\begin{align}
    &
    K = 
    \left(
    \begin{array}{cccc|c}
        -1 & 2 & 2 & -1 & 3 \\
        2 & 0 & -1 & 0 & -3 \\
        2 & -1 & -1 & 2 & -3 \\
        -1 &  0 & 2 & 0 & 3 \\
        \hline
        3 & -3 & -3 & 3 & 0
    \end{array}
    \right)
    \;,\;\;
    C = 
    \left(
    \begin{array}{cccc}
        1 & 0 & 0 & 0 \\
        0 & 1 & 0 & 0 \\
        0 & 0 & 1 & 0 \\
        0 & 0 & 0 & 1 \\
        \hline
        0 & 0 & 0 & 0
    \end{array}
    \right)
    \;,
    \nonumber\\
    &\CV =
    \{
    \phi_2^2 V_{(-1,0,0,1)} ,
    \phi_1\phi_3^2 V_{(0,-1,0,-1)} ,
    \phi_4^2 V_{(0,1,-1,0)} 
    \}
    \,,
\end{align}
where the unbroken topological symmetry is generated by $A = T_1 - T_2 - T_3 + T_4$. By turning off the Higgs branch symmetry, we can parametrize the mixing as $\m = (\n, -3 -\n, -4 -\n, 1 + \n)$. The F-maximization computation exhibits a flat direction along $\n$, suggesting that $U(1)_A$ decouples in the IR with free energy $F = \log \sqrt{8}$, or equivalently $|S_{00}| = 1/\sqrt{8}$. However, the values of the handle gluings $\CH_\a$ vary as we vary $\n$, which appears to be in tension with this flat direction. We leave a detailed understanding of this behavior for future work. For the purposes of extracting the modular data, we set $\n=n/2$ for odd integer $n$, with which the resulting TQFT is unitary.

The simple lines and modular data are
\begin{align}
\begin{split}
    \L =& \{1 ,\;z_1z_2^{-1}z_3^{-1}z_4,\; z_1^2z_2^{-2}z_3^{-2}z_4^2,\; z_1^3z_2^{-3}z_3^{-3}z_4^3,\;z_1^2 z_2^{-2}z_3^{-1}z_4^2+z_1z_2^{-1} z_3^{-1}z_4,\;1+z_1z_2^{-1}z_4\},\\\\
    \;\to&\;
    S=
    \frac{1}{2\sqrt{2}}
\begin{pmatrix}
1 & 1 & 1 & 1 & \sqrt{2} & \sqrt{2} \\
1 & -1 & 1 & -1 & -\sqrt{2}\,i & \sqrt{2}\,i \\
1 & 1 & 1 & 1 & -\sqrt{2} & -\sqrt{2} \\
1 & -1 & 1 & -1 & \sqrt{2}\,i & -\sqrt{2}\,i \\
\sqrt{2} & -\sqrt{2}\,i & -\sqrt{2} & \sqrt{2}\,i & 0 & 0 \\
\sqrt{2} & \sqrt{2}\,i & -\sqrt{2} & -\sqrt{2}\,i & 0 & 0
\end{pmatrix}
    ,\;
    h=\Big(0,\frac14,\frac12,\frac14,\frac58,\frac58\Big)
    .\;
\end{split}
\label{eq: A5 ST}
\end{align}
The result for $S$ agrees up to an overall factor of $4i$ and signed permutation of basis with that found for the atypical modules of the $\CB_4$-algebra in \cite{Auger:2019gts}.  Further, using a different approach  \cite{adamovic2019classificationirreduciblemodulesbershadskypolyakov} found results that indicate the existence of a rank 6 category of distinguished modules in the $\CB_4$-algebra with conformal weights
$$
h=\left(0,-\frac14,-\frac12,-\frac14,-\frac38,-\frac38\right).
$$
Our result for the topological spins in this case agrees with this mod $\frac12$ as expected.  Together, these are a nontrivial check on applying our algorithm to the scenario where the IR TQFT is spin.

\subsubsection{\texorpdfstring{$A_6$}{A6} theory}
The monodromy operator of $A_6$ theory has periodicity $p=9$ so we only consider the coprime cases $n=1,2,4$. For each power we find a numerical instability in our calculations. 
%This is a common trait of theories with modular $S$-matrices containing $0$'s. 
%
This instability is mitigated either by moving slightly away from the SCFT point along the direction of the unbroken generator $A$, or by beginning with a proper subset of the available monopole operators in the superpotential and taking appropriate limits in the parameter space of IR R-symmetries to turn on the remaining monopoles.  When the latter is necessary we specify the removed operators.
\paragraph{\texorpdfstring{$\boldsymbol{{\rm TFT}[A_6^{(1),\n}]}$}{A6(1)}}
The monodromy trace $\Tr \Phi$ gives rise to a 3d $\CN=2$ $U(1)^3$ ACSM theory
\begin{align}
    K = 
    \begin{pmatrix}
        1 & -2 & 0 \\
        -2 & 3 & 2 \\
        0 & 2 & 2
    \end{pmatrix}
    \;,\;
    C = I_3
    \;,\;
    \CV = \{
    \phi_1^2 V_{(0,1,-1)}
    ,
    \phi_2^2 V_{(1,0,0)}
    \}
    \,,
\end{align}
which flows to a rank-0 SCFT with the $\CN=4$ axial symmetry generated by $A = T_2 + T_3$.  We find the simple lines and the modular data at the 3d A-/B-twist sectors ${\rm TFT}[A_6^{(1),\n}]$ to be
\begin{align}
    &\n=-
    \;:\;
    \L = \{ 1 , -z_2^{-1}, -z_2^{-1}z_3^{-1},z_1^{-1}z_2 z_3 \}
    \nonumber\\
    &\qquad\to\quad
    S=
    \frac{2}{3}
    \begin{pmatrix}
        \sin \frac{2\pi}{9} & \sin \frac{3\pi}{9} & -\sin \frac{\pi}{9} & -\sin \frac{4\pi}{9} \\
        \sin \frac{3\pi}{9} & 0 & -\sin \frac{3\pi}{9} & \sin \frac{3\pi}{9} \\
        -\sin\frac{\pi}{9} & -\sin\frac{3\pi}{9} & -\sin\frac{4\pi}{9} & -\sin \frac{2\pi}{9} \\
        -\sin \frac{4\pi}{9} & \sin \frac{3\pi}{9} & -\sin \frac{2\pi}{9} & \sin \frac{\pi}{9}
    \end{pmatrix}
    ,\;
    h=\Big(0,\frac{4}{9},\frac{1}{3},\frac{2}{3}\Big)
    ,\;
    c=\frac{14}{3}   
    \,.
    \label{eq: A6(1)A}
    \\
    &\n=+
    \;:\;
    \L = \{ 1 , -z_1^{-1} z_2^2 z_3^2 , z_2 z_3 , -z_3 \}
    \nonumber\\
    &\qquad\to\quad
    S=
    \frac{2}{3}
    \begin{pmatrix}
        \sin \frac{4\pi}{9} & \sin \frac{2\pi}{9} & -\sin \frac{\pi}{9} & -\sin \frac{3\pi}{9} \\
        \sin \frac{2\pi}{9} & -\sin\frac{\pi}{9} & -\sin \frac{4\pi}{9} & \sin \frac{3\pi}{9} \\
        -\sin\frac{\pi}{9} & -\sin\frac{4\pi}{9} & -\sin\frac{2\pi}{9} & -\sin \frac{3\pi}{9} \\
        -\sin \frac{3\pi}{9} & \sin \frac{3\pi}{9} & -\sin \frac{3\pi}{9} & 0
    \end{pmatrix}
    ,\;
    h=\Big(0,\frac{1}{3},\frac{2}{3},\frac{1}{9}\Big)
    ,\;
    c=\frac{2}{3}
    \,.
    \label{eq: A6(1)B}
\end{align}
The modular labels are $4^{9,544}_{\frac{14}{3},5.445}$ and $4^{9,199}_{\frac{2}{3},2.319}$. The modular data \eqref{eq: A6(1)A} are compatible with $M(2,9)$ with $c=-46/3$ upon taking overall sign flip on $S$, while \eqref{eq: A6(1)B} is the one from affine $osp(1|2)$ at level 3.

\paragraph{\texorpdfstring{$\boldsymbol{{\rm TFT}[A_6^{(2),\n}]}$}{A6(2)}}
The quadratic power trace $\Tr \Phi^2$ gives rise to a 3d $\CN=2$ $U(1)^3$ ACSM theory
\begin{align}
    K = 
    \begin{pmatrix}
        2 & -2 & 0 \\
        -2 & 2 & 2 \\
        0 & 2 & 0
    \end{pmatrix}
    \;,\;
    C = I_3
    \;,\;
    \CV = \{
    \phi_2^2 V_{(0,0,-1)}
    ,
    \phi_3^2 V_{(-1,-1,0)}
    \}
    \,,
\end{align}
which flows to a rank-0 SCFT with the $\CN=4$ axial symmetry generated by $A = T_1 - T_2$.  The simple lines and the modular data at the 3d A-/B-twist sectors ${\rm TFT}[A_6^{(2),\n}]$ are identified
\begin{align}
    &\n=-
    \;:\;
    \L = \{ 1 , -z_1^{-2}z_2^2 z_3 , -z_1^2 z_2^{-1}, z_1^{-1}z_2^2 \}
    \nonumber\\
    &\qquad\to\quad
    S=
    \frac{2}{3}
    \begin{pmatrix}
        \sin \frac{4\pi}{9} & \sin \frac{2\pi}{9} & -\sin \frac{\pi}{9} & -\sin \frac{3\pi}{9} \\
        \sin \frac{2\pi}{9} & -\sin \frac{\pi}{9} & -\sin \frac{4\pi}{9} & \sin \frac{3\pi}{9} \\
        -\sin\frac{\pi}{9} & -\sin\frac{4\pi}{9} & -\sin\frac{2\pi}{9} & -\sin \frac{3\pi}{9} \\
        -\sin \frac{3\pi}{9} & \sin \frac{3\pi}{9} & -\sin \frac{3\pi}{9} & 0
    \end{pmatrix}
    ,\;
    h=\Big(0,\frac{2}{3},\frac{1}{3},\frac{8}{9}\Big)
    ,\;
    c=\frac{22}{3}   
    \,.
    \label{eq: A6(2)A}
    \\
    &\n=+
    \;:\;
    \L = \{ 1 , -z_2, z_1 z_2^{-1}, -z_1^{-1}z_2 z_3\}
    \nonumber\\
    &\qquad\to\quad
    S=
    \frac{2}{3}
    \begin{pmatrix}
        \sin \frac{2\pi}{9} & \sin \frac{3\pi}{9} & -\sin \frac{\pi}{9} & -\sin \frac{4\pi}{9} \\
        \sin \frac{3\pi}{9} & 0 & -\sin \frac{3\pi}{9} & \sin \frac{3\pi}{9} \\
        -\sin\frac{\pi}{9} & -\sin\frac{3\pi}{9} & -\sin\frac{4\pi}{9} & -\sin \frac{2\pi}{9} \\
        -\sin \frac{4\pi}{9} & \sin \frac{3\pi}{9} & -\sin \frac{2\pi}{9} & \sin \frac{\pi}{9}
    \end{pmatrix}
    ,\;
    h=\Big(0,\frac{5}{9},\frac{2}{3},\frac{1}{3}\Big)
    ,\;
    c=\frac{10}{3}
    \,.
    \label{eq: A6(2)B}
\end{align}
The modular labels are 
$4^{9,549}_{\frac{22}{3},2.319}$ and $4^{9,616}_{\frac{10}{3},5.445}$. The modular data \eqref{eq: A6(2)A} and \eqref{eq: A6(2)B} are complex conjugate of \eqref{eq: A6(1)B} and \eqref{eq: A6(1)A} respectively.

\paragraph{\texorpdfstring{$\boldsymbol{{\rm TFT}[A_6^{(4)}]}$}{A6(4)}}
The monodromy trace $\Tr \Phi^4$ gives a 3d $\CN=2$ ACSM theory as
\footnote{
To find the correct four Bethe vacua, we first solve the vacuum equation without the second term in $\CV$ then tune the unbroken real mass parameter to the value determined by turning on the second term.
}
\begin{align}
    &\qquad\qquad
    K = 
    \begin{pmatrix}
        0 & 2 & 0 & 0 & 0 & 0 \\
        2 & 0 & -2 & 2 & 0 & 0 \\
        0 & -2 & 0 & 0 & 2 & 0 \\
        0 & 2 & 0 & 0 & -2 & 2 \\
        0 & 0 & 2 & -2 & 0 & 0 \\
        0 & 0 & 0 & 2 & 0 & 0
    \end{pmatrix}
    ,\;
    C = I_6
    ,
    \nonumber\\
    &\CV
    =
    \{
    \phi_1^2 V_{(0,-1,0,0,-1,0)},
    \phi_2^2 V_{(-1,0,0,0,0,0)},
    \phi_3^2 V_{(0,0,0,0,-1,-1)},
    \nonumber\\
    &\qquad\qquad
    \phi_4^2 V_{(0,0,0,0,0,-1)},
    \phi_5^2 V_{(-1,0,-1,0,0,0)},
    \phi_6^2 V_{(0,0,-1,-1,0,0)}
    \}
    \,.
\end{align}
This theory flows to a unitary TQFT ${\rm TFT}[A_6^{(4)}]$ with simple lines and modular data
\begin{align}
    &\L = \{
    1 , -z_1 z_3^{-1}z_4 , - z_4 , z_2^{-1} z_5
    \}
    \nonumber\\
    &\qquad
    \to\;\;
    S=
    \frac{2}{3}
    \begin{pmatrix}
        \sin \frac{\pi}{9} & \sin \frac{2\pi}{9} & \sin \frac{3\pi}{9} & \sin \frac{4\pi}{9} \\
        \sin \frac{2\pi}{9} & -\sin \frac{4\pi}{9} & \sin \frac{3\pi}{9} & -\sin \frac{\pi}{9} \\
        \sin\frac{3\pi}{9} & \sin\frac{3\pi}{9} & 0 & -\sin \frac{3\pi}{9} \\
        \sin \frac{4\pi}{9} & -\sin \frac{\pi}{9} & -\sin \frac{3\pi}{9} & \sin \frac{2\pi}{9}
    \end{pmatrix}
    ,\;
    h
    =\Big( 0,\frac{2}{3},\frac{7}{9},\frac{1}{3} \Big)
    ,\;
    c=\frac{14}{3}
\end{align}
The modular label is $4^{9,614}_{\frac{14}{3},19.23}$.

\subsubsection{\texorpdfstring{$D_4$}{D4} theory}
\paragraph{\texorpdfstring{$\boldsymbol{{\rm TFT}[D_4^{(1)}]}$}{D4(1)}}
The 3d $\CN=2$ gauge theory arising from the trace $\Tr \Phi$ of the $D_4$ theory becomes a pure $U(1)^2 \times U(1)_{f_1} \times U(1)_{f_2} $ CS theory where $U(1)_{f_i}$'s correspond to the Cartan of the SU(3) flavor symmetry in the theory
\begin{align}
    K = 
    \left(
    \begin{array}{cc|cc}
        0 & -2 & 1 & -1 \\
        -2 & 0 & 1 & 0 \\
        \hline
        1 & 1 & 2 & 2 \\
        -1 & 0 & 2 & 3
    \end{array}
    \right)
    \,.
\end{align}
This theory flows to a unitary TQFT ${\rm TFT}[D_4^{(1)}]$ with four consistent simple line choices
\begin{align}
    &\L^{(1)} = \{ 1 , z_1 , z_2, z_1 z_2 \}
    \;\;,\;\;
    \L^{(2)} = \{ 1 , -z_1 , -z_2, z_1 z_2 \}
    \;,
    \nonumber\\
    &\L^{(3)} = \{ 1 , z_1 , -z_2, -z_1 z_2 \}
    \;\;,\;\;
    \L^{(4)} = \{ 1 , -z_1 , z_2, -z_1 z_2 \}
    \,,
\end{align}
which give rise to five possible consistent modular data
\begin{align}
    &\L^{(1)} \;\;\to\;\; \big\{ S^{(1)},h^{(1)},c^{(1)} \big\}
    \,,
    \nonumber\\
    &\L^{(2)} \;\;\to\;\; \big\{ S^{(1)},h^{(3)},c^{(2)} \big\}
    \;,\;
    \big\{ S^{(2)},h^{(2)},c^{(1)} \big\}
    \,,
    \nonumber\\
    &\L^{(3)} \;\;\to\;\; \big\{ S^{(2)},h^{(1)},c^{(1)} \big\}
    \,,
    \nonumber\\
    &\L^{(4)} \;\;\to\;\; \big\{ S^{(2)},h^{(3)},c^{(2)} \big\}
    \,,
    \label{eq: D4 modular data}
\end{align}
where
\begin{align*}
\begin{split}
S^{(1)}&=
\frac{1}{2}
\begin{pmatrix}
1 & 1 & 1 & 1\\
1 & 1 & -1 & -1\\
1 & -1 & 1 & -1\\
1 & -1 & -1 & 1
\end{pmatrix},\;\;
S^{(2)} =
\frac{1}{2}
\begin{pmatrix}
1 & 1 & -1 & -1\\
1 & 1 & 1 & 1\\
-1 & 1 & 1 & -1\\
-1 & 1 & -1 & 1
\end{pmatrix},
\end{split}
\\
\\
\begin{split}
    h^{(1)} = \left(0,0,0,\frac12\right),\;\;\;
    &h^{(2)} = \left(0,\frac12,0,0\right),\;\;\;
    h^{(3)} = \left(0,\frac12,\frac12,\frac12\right),\;\;\;
    c^{(1)} = 0,\;\;\;
    c^{(2)} = 4.\;\;\;
\end{split}
\end{align*}
None of them are compatible with affine $su(3)$ at admissible level $-3/2$~\cite{DiFrancesco:1997nk}
\begin{align}
    S = \frac{1}{2}
    \begin{pmatrix}
        -1 & -1 & -1 & 1 \\
        -1 & -1 & 1 & -1 \\
        -1 & 1 & -1 & -1 \\
        1 & -1 & -1 & -1
    \end{pmatrix}
    \;,\;\;
    h=\Big( 0, \frac{1}{2}, \frac{1}{2}, \frac{1}{2} \Big)
    \;,\;\;
    c=-8
    \,,
\end{align}
since this involves negative fusion coefficients that is ruled out from our routine. 

The modular labels, reading from top to bottom, left to right in \eqref{eq: D4 modular data} are: $4^{2,750}_{0,4}$, $4^{2,250}_{4,4}$, $4^{2,250}_{0,4}$, $4^{2,250}_{0,4}$, $4^{2,250}_{4,4}$ where the first two are unitary modular data.  Note there are collisions in labeling here.  This occurs because the modular data are distinguished by whether or not a spin-1/2 object gets a quantum dimension of $+1$ or $-1$, a possibility the modular label convention cannot distinguish.  Nonetheless they furnish distinct modular data which can be found in the final appendix of \cite{ng2025classificationmodulardatarank}.  All of these modular data lie in Galois orbits of order 1 as expected from the periodicity of this case.

\subsubsection{\texorpdfstring{$D_5$}{D5} theory}
The $D_5$ theory has periodicity $p=5$ so we consider $\Tr \Phi^n$ with $n=1,2$.
\paragraph{\texorpdfstring{$\boldsymbol{{\rm TFT}[D_5^{(1)}]}$}{D5(1)}}
The monodromy trace $\Tr \Phi$ gives rise to a 3d $\CN=2$ pure $U(1)\times U(1)_f$ CS theory where $U(1)_f$ corresponds to Cartan of the $SU(2)$ flavor symmetry in $D_5$ theory
\begin{align}
    K
    =
    \left(
    \begin{array}{c|c}
        5 & 1 \\
    \hline
        1 & 1
    \end{array}
    \right)
    \,,
\end{align}
which flows to a unitary TQFT ${\rm TFT}[D_5^{(1)}]$. We identify the simple lines and modular data to be ($\z = e^{\frac{2\pi i}{5}}$)
\begin{align}
    &\L = \{ 1 , -z_1^3 , z_1^2 , z_1^4 , -z_1 \}
    \nonumber\\
    &\qquad\to\;\;
    S = \frac{1}{\sqrt{5}}
    \begin{pmatrix}
        1 & 1 & 1 & 1 & 1 \\
        1 & \z & \z^4 & \z^3 & \z^2 \\
        1 & \z^4 & \z & \z^2 & \z^3 \\
        1 & \z^3 & \z^2 & \z^4 & \z \\
        1 & \z^2 & \z^3 & \z & \z^4 \\
    \end{pmatrix}
    \;,\;
    h = \Big(
    0 , \frac{2}{5} , \frac{2}{5} , \frac{3}{5} , \frac{3}{5}
    \Big)
    \;,\;
    c=4
    \,.
    \label{eq: D5(1)}
\end{align}
The modular label is $5^{5,210}_{4,5}$. This is different from the modular data of affine $su(2)$ at level $-8/5$ with $c=-12$ which has negative fusion coefficients, and thus, ruled out from our procedure.

\paragraph{\texorpdfstring{$\boldsymbol{{\rm TFT}[D_5^{(2)}]}$}{D5(2)}}
The trace of the quadratic power of the monodromy $\Tr \Phi^2$ reads a 3d $\CN=2$ pure $U(1)^2\times U(1)_f$ CS theory
\begin{align}
    K
    =
    \left(
    \begin{array}{cc|c}
        3 & -4 & -2 \\
        -4 & 7 & 3 \\
        \hline
        -2 & 3 & 3
    \end{array}
    \right)
    \,,
\end{align}
which flows to a unitary TQFT ${\rm TFT}[D_5^{(2)}]$. We find the simple lines and modular data to be
\begin{align}
    &\L = \{ 1 , -z_1 , -z_1^2 z_2 , z_1 z_2 , -z_2 \}
    \nonumber\\
    &\qquad\to\;\;
    S = \frac{1}{\sqrt{5}}
    \begin{pmatrix}
        1 & 1 & 1 & 1 & 1 \\
        1 & \z^3 & \z^2 & \z^4 & \z \\
        1 & \z^2 & \z^3 & \z & \z^4 \\
        1 & \z^4 & \z & \z^2 & \z^3 \\
        1 & \z & \z^4 & \z^3 & \z^2 \\
    \end{pmatrix}
    \;,\;
    h = \Big(
    0 , \frac{1}{5} , \frac{1}{5} , \frac{4}{5} , \frac{4}{5}
    \Big)
    \;,\;
    c=0
    \,.
\end{align}
The modular label is $5^{5,110}_{0,5}$. Observe that these modular data can be obtained by taking $\z \to \z^2$ and $h \to 2h \mod 1$ in \eqref{eq: D5(1)}, i.e., Galois conjugation as expected in~\cite{Go:2025ixu}.

\subsubsection{\texorpdfstring{$D_6$}{D6} theory}
Since $\Phi^3 \sim {\bf 1}$ for the $D_6$ theory, we only consider $\Tr \Phi$.
\paragraph{\texorpdfstring{$\boldsymbol{{\rm TFT}[D_6^{(1)}]}$}{D6(1)}}
The monodromy trace $\Tr \Phi$ gives a 3d $\CN=2$ $U(1)^2 \times U(1)_{f_1} \times U(1)_{f_2}$ pure CS theory
\begin{align}
    K = 
    \left(
    \begin{array}{cc|cc}
        -3 & -3 & 0 & -1 \\
        -3 & 0 & 0 & -1 \\
        \hline
        0 & 0 & 1 & 0 \\
        -1 & -1 & 0 & 3
    \end{array}
    \right)
    \,,
\end{align}
and we identify the simple lines and modular data to be ($\omega = e^{\frac{2\pi i}{3}}$)
\begin{align}
    &\L = \{ 
    1 , 
    z_1 ,
    z_1^2 ,
    z_1 z_2^2 ,
    -z_1^2 z_2 ,
    z_1^2 z_2^2 ,
    -z_1 z_2 ,
    z_2^2 ,
    -z_2
    \}
    \nonumber\\
    &\to\;\;
    S = \frac{1}{3}
    \begin{pmatrix}
        1 & 1 & 1 & 1 & 1 & 1 & 1 & 1 & 1\\
        1 & 1 & 1 & \omega^2 & \omega & \omega^2 & \omega & \omega^2 & \omega\\
        1 & 1 & 1 & \omega & \omega^2 & \omega & \omega^2 & \omega & \omega^2\\
        1 & \omega^2 & \omega & 1 & 1 & \omega^2 & \omega & \omega & \omega^2\\
        1 & \omega & \omega^2 & 1 & 1 & \omega & \omega^2 & \omega^2 & \omega\\
        1 & \omega^2 & \omega & \omega^2 & \omega & \omega & \omega^2 & 1 & 1\\
        1 & \omega & \omega^2 & \omega & \omega^2 & \omega^2 & \omega & 1 & 1\\
        1 & \omega^2 & \omega & \omega & \omega^2 & 1 & 1 & \omega^2 & \omega\\
        1 & \omega & \omega^2 & \omega^2 & \omega & 1 & 1 & \omega & \omega^2
    \end{pmatrix}
    \;,\;
    h = \Big(
    0 , 0 , 0 , 0 , 0 , \frac{1}{3}, \frac{1}{3}, \frac{2}{3}, \frac{2}{3}
    \Big)
    \;,\;
    c=0
    \,.
\end{align}
The modular label is $9^{3,113}_{0,9}$.

\subsubsection{\texorpdfstring{$E_6$}{E6} theory}
The monodromy operator of $E_6$ theory has periodicity $7$, thereby, we consider $\Tr \Phi^n$ for $n=1,2,3$.
\paragraph{\texorpdfstring{$\boldsymbol{{\rm TFT}[E_6^{(1),\n}]}$}{E6(1)}}
The 3d $\CN=2$ UV gauge theory from $\Tr \Phi$ is 
\begin{align}
    K
    =
    \begin{pmatrix}
        2 & 1 & -1 \\
        1 & 1 & 1 \\
        -1 & 1 & 2
    \end{pmatrix}
    \;,\;
    C = I_3 
    \;,\;
    \CV = 
    \{
    \phi_1^2 V_{(0,-1,1)}
    ,
    \phi_3^2 V_{(1,-1,0)}
    \}
    \,,
\end{align}
which flows to a rank-0 SCFT with the $\CN=4$ axial symmetry generated by $A = T_1 + T_2 + T_3$. The 3d topological A/B-twists become non-unitary topological field theories ${\rm TFT}[E_6^{(1),\n}]$ with simple lines and modular data are identified to be
\begin{align}
    &
    \n = - \;:\;
    \L = \{
    1 , z_1^{-1}+ z_1 z_2 , z_1^{-1}z_2^{-1}z_3^{-1}, z_1^{-1}, z_3^{-1}
    \}
    \nonumber\\
    &\to\;\;
    S = \frac{2}{\sqrt{7}}
    \begin{pmatrix}
    \sin\frac{3\pi}{14} &
    \sin\frac{5\pi}{14} &
    -\sin\frac{\pi}{14} &
    -\sin\frac{\pi}{6} &
    -\sin\frac{\pi}{6}
    \\
    \sin\frac{5\pi}{14} &
    -\sin\frac{\pi}{14} &
    -\sin\frac{3\pi}{14} &
    \sin\frac{\pi}{6} &
    \sin\frac{\pi}{6}
    \\
    -\sin\frac{\pi}{14} &
    -\sin\frac{3\pi}{14} &
    -\sin\frac{5\pi}{14} &
    -\sin\frac{\pi}{6} &
    -\sin\frac{\pi}{6}
    \\
    -\sin\frac{\pi}{6} &
    \sin\frac{\pi}{6} &
    -\sin\frac{\pi}{6} &
    \frac{1-\sqrt{7}i}{4} &
    \frac{1+\sqrt{7}i}{4}
    \\
    -\sin\frac{\pi}{6} &
    \sin\frac{\pi}{6} &
    -\sin\frac{\pi}{6} &
    \frac{1+\sqrt{7}i}{4} &
    \frac{1-\sqrt{7}i}{4}
    \end{pmatrix}
    \;,\;
    h = \Big(
    0,\frac{3}{7},\frac{2}{7},\frac{4}{7},\frac{4}{7}
    \Big)
    \;,\;
    c=\frac{26}{7}
    \,.
    \label{eq: E6(1)A}
    \\
    &
    \n = + \;:\;
    \L = \{
    1 , z_2 z_3, z_1 z_2, -z_1 z_2 z_3 , -1 + z_1 z_2^2 z_3
    \}
    \nonumber\\
    &\to\;\;
    S = \frac{2}{\sqrt{7}}
    \begin{pmatrix}
    \sin\frac{5\pi}{14} &
    \sin\frac{\pi}{6} &
    \sin\frac{\pi}{6} &
    -\sin\frac{\pi}{14} &
    -\sin\frac{3\pi}{14}
    \\
    \sin\frac{\pi}{6} &
    \frac{-1+\sqrt{7}i}{4} &
    \frac{-1-\sqrt{7}i}{4}  &
    -\sin\frac{\pi}{6} &
    \sin\frac{\pi}{6}
    \\
    \sin\frac{\pi}{6} &
    \frac{-1-\sqrt{7}i}{4} &
    \frac{-1+\sqrt{7}i}{4} &
    -\sin\frac{\pi}{6} &
    \sin\frac{\pi}{6}
    \\
    -\sin\frac{\pi}{14} &
    -\sin\frac{\pi}{6} &
    -\sin\frac{\pi}{6} &
    -\sin\frac{3\pi}{14} &
    -\sin\frac{5\pi}{14}
    \\
    -\sin\frac{3\pi}{14} &
    \sin\frac{\pi}{6} &
    \sin\frac{\pi}{6} &
    -\sin\frac{5\pi}{14} &
    \sin\frac{\pi}{14}
    \end{pmatrix}
    \;,\;
    h = \Big(
    0,\frac{2}{7},\frac{2}{7},\frac{5}{7},\frac{1}{7}
    \Big)
    \;,\;
    c=\frac{6}{7}
    \,.
    \label{eq: E6(1)B}
\end{align}
The modular labels are $5^{7,408}_{\frac{26}{7},4.501}$ and $5^{7,342}_{\frac{6}{7},2.155}$ respectively. Note that each $\L$ contains a multinomial for which we could not identify the simple line as a Wilson loop; see appendix \ref{app: A-model} for computations relevant to this. The computed modular data \eqref{eq: E6(1)A} are that of the desired VOA, the $W_3(3,7)$ minimal model~\cite{Gannon:1992ty,Beltaos:2010ka} up to flipping an overall sign of $S$.

\paragraph{\texorpdfstring{$\boldsymbol{{\rm TFT}[E_6^{(2)}]}$}{E6(2)}}
The trace of quadratic monodromy gives rise to a 3d $\CN=2$ ACSM theory as
\begin{align}
    &\qquad\qquad\quad
    K
    =
    \begin{pmatrix}
        0 & -1 & -1 & 2 & 2 & 0\\
        -1 & 0 & -1 & 2 & 0 & 2 \\
        -1 &-1 & -1 & 2 & 2 & 2 \\
        2 & 2 & 2 & -1 & -1 & -1 \\
        2 & 0 & 2 & -1 & 0 & -1 \\
        0 & 2 & 2 & -1 & -1 & 0
    \end{pmatrix}
    ,\;
    C = I_6
    ,\;
    \nonumber\\
    &
    \CV=
    \{
    \phi_1^2 V_{(0,0,0,-1,0,1)},
    \phi_2^2 V_{(0,0,0,-1,1,0)},
    \phi_3^2 V_{(0,0,0,1,-1,-1)},
    \nonumber\\
    &\qquad\qquad\qquad\qquad
    \phi_4^2 V_{(-1,-1,1,0,0,0)},
    \phi_5^2 V_{(0,1,-1,0,0,0)},
    \phi_6^2 V_{(1,0,-1,0,0,0)}
    \}
\end{align}
which flows to a unitary topological field theory ${\rm}TFT[E_6^{(2)}]$ with simple lines and modular data are identified to be
\begin{align}
    &\L = \{ 
    1 , z_2 z_3, z_4 z_5 , - z_4 z_5 z_6 , -1 + z_2 z_3 z_4 z_5
    \}
    \nonumber\\
    &\to\;\;
    S = \frac{2}{\sqrt{7}}
    \begin{pmatrix}
    \sin\frac{\pi}{14} &
    \sin\frac{\pi}{6} &
    \sin\frac{\pi}{6} &
    \sin\frac{3\pi}{14} &
    \sin\frac{5\pi}{14}
    \\
    \sin\frac{\pi}{6} &
    \frac{-1+\sqrt{7}i}{4} &
    \frac{-1-\sqrt{7}i}{4} &
    -\sin\frac{\pi}{6} &
    \sin\frac{\pi}{6}
    \\
    \sin\frac{\pi}{6} &
    \frac{-1-\sqrt{7}i}{4}&
    \frac{-1+\sqrt{7}i}{4} &
    -\sin\frac{\pi}{6} &
    \sin\frac{\pi}{6}
    \\
    \sin\frac{3\pi}{14} &
    -\sin\frac{\pi}{6} &
    -\sin\frac{\pi}{6} &
    \sin\frac{5\pi}{14} &
    -\sin\frac{\pi}{14}
    \\
    \sin\frac{5\pi}{14} &
    \sin\frac{\pi}{6} &
    \sin\frac{\pi}{6} &
    -\sin\frac{\pi}{14} &
    -\sin\frac{3\pi}{14}
    \end{pmatrix}
    \;,\;
    h = \Big(
    0,\frac{1}{7},\frac{1}{7},\frac{6}{7},\frac{4}{7}
    \Big)
    \;,\;
    c=\frac{38}{7}
    \,.
\end{align}
The modular label is $5^{7,386}_{\frac{38}{7},35.34}$

\paragraph{\texorpdfstring{$\boldsymbol{{\rm TFT}[E_6^{(3),\n}]}$}{E6(3)}}
The trace of cubic monodromy reads a 3d $\CN=2$ ACSM theory
\begin{align}
    K
    =
    \begin{pmatrix}
        1 & -1 & -1 & 2 \\
        -1 & 1 & 1 & 0 \\
        -1 & 1 & 0 & 0 \\
        2 & 0 & 0 & 0
    \end{pmatrix}
    ,\;
    C = I_4
    ,\;
    \CV=
    \{
    \phi_1^2 V_{(0,0,0,-1)},
    \phi_4^2 V_{(-1,-1,0,0)},
    \phi_1 \phi_2 V_{(0,0,-1,-1)}
    \}
    \,,
\end{align}
which flows to a rank-0 SCFT point with $\CN=4$ axial symmetry generated by $A = -T_1 + T_2$. 

We note that there are two possible choices of superpotential terms for this theory, and that the one we use here is not the same as that of~\cite{Go:2025ixu}.  Initially the two choices appear to be equivalent as they lead to the same fixed point.  Indeed, one can check that the superconformal indices agree.  However, the partial modular data $S_{0\a}$ computed in~\cite{Go:2025ixu} could not lie in the Galois orbit of the other $E_6$ theories.  The reason for this discrepancy is as follows. Like the $A_6$ cases, this theory is numerically unstable and one must remove a monopole operator and limit toward the fixed point in parameter space\footnote{
We solve the vacuum equation without the first term in $\CV$ to obtain six vacua, then tune the remaining real mass parameter as fixed by the first term. Three $S_{0\a}$ vanishes and we discard the corresponding vacua to get the modular data.}.  The choice of superpotential determines the direction for this limit by requiring the R-charge remain 2.  We find that the resulting S-matrix depends on this direction and that our choice of superpotential admits the expected Galois conjugate of the modular data.

The 3d topological A-/B-twists capture non-unitary topological theories ${\rm TFT}[E_6^{(3),\n}]$ with simple lines and modular data
\begin{align}
    &
    \n = - \;:\;
    \L = \{
    1 , -z_1 -z_3^{-1}z_4 , z_1 z_2^{-1}z_3^{-1}z_4 , -z_1 z_2^{-1}, z_1^2 z_2^{-1}
    \}
    \nonumber\\
    &\to\;\;
    S = \frac{2}{\sqrt{7}}
    \begin{pmatrix}
    \sin\frac{5\pi}{14} &
    \sin\frac{\pi}{6} &
    \sin\frac{\pi}{6} &
    -\sin\frac{\pi}{14} &
    -\sin\frac{3\pi}{14}
    \\
    \sin\frac{\pi}{6} &
    \frac{-1-\sqrt{7}i}{4} &
    \frac{-1+\sqrt{7}i}{4} &
    -\sin\frac{\pi}{6} &
    \sin\frac{\pi}{6}
    \\
    \sin\frac{\pi}{6} &
    \frac{-1+\sqrt{7}i}{4} &
    \frac{-1-\sqrt{7}i}{4} &
    -\sin\frac{\pi}{6} &
    \sin\frac{\pi}{6}
    \\
    -\sin\frac{\pi}{14} &
    -\sin\frac{\pi}{6} &
    -\sin\frac{\pi}{6} &
    -\sin\frac{3\pi}{14} &
    -\sin\frac{5\pi}{14}
    \\
    -\sin\frac{3\pi}{14} &
    \sin\frac{\pi}{6} &
    \sin\frac{\pi}{6} &
    -\sin\frac{5\pi}{14} &
    \sin\frac{\pi}{14}
    \end{pmatrix}
    \;,\;
    h = \Big(
    0,\frac{5}{7},\frac{5}{7},\frac{2}{7},\frac{6}{7}
    \Big)
    ,\;
    c=\frac{50}{7}
    \,.
    \label{eq: E6(3)A}
    \\
    &
    \n = + \;:\;
    \L = \{
    1 , -z_1 , z_1^{-1}z_2 , z_3^{-1}z_4, -z_2 - z_1^{-1}z_2 z_3^{-1}z_4
    \}
    \nonumber\\
    &\to\;\;
    S = \frac{2}{\sqrt{7}}
    \begin{pmatrix}
    \sin\frac{3\pi}{14} &
    \sin\frac{5\pi}{14} &
    -\sin\frac{\pi}{14} &
    -\sin\frac{\pi}{6} &
    -\sin\frac{\pi}{6}
    \\
    \sin\frac{5\pi}{14} &
    -\sin\frac{\pi}{14} &
    -\sin\frac{3\pi}{14} &
    \sin\frac{\pi}{6} &
    \sin\frac{\pi}{6}
    \\
    -\sin\frac{\pi}{14} &
    -\sin\frac{3\pi}{14} &
    -\sin\frac{5\pi}{14} &
    -\sin\frac{\pi}{6} &
    -\sin\frac{\pi}{6}
    \\
    -\sin\frac{\pi}{6} &
    \sin\frac{\pi}{6} &
    -\sin\frac{\pi}{6} &
    \frac{1+\sqrt{7}i}{4} &
    \frac{1-\sqrt{7}i}{4}
    \\
    -\sin\frac{\pi}{6} &
    \sin\frac{\pi}{6} &
    -\sin\frac{\pi}{6} &
    \frac{1-\sqrt{7}i}{4} &
    \frac{1+\sqrt{7}i}{4}
    \end{pmatrix}
    \;,\;
    h = \Big(
    0,\frac{4}{7},\frac{5}{7},\frac{3}{7},\frac{3}{7}
    \Big)
    \;,\;
    c=\frac{30}{7}
    \,.
    \label{eq: E6(3)B}
\end{align}
The modular labels are $5^{7,255}_{\frac{50}{7},2.155}$ and $5^{7,125}_{\frac{30}{7},4.501}$ respectively.

\section*{Acknowledgements}
We would like to thank Heeyeon Kim for early contributions and valuable discussions. We also thank Anindya Banerjee, Cyril Closset, Leonardo Rastelli, Ranveer Singh, and Madhav Sinha for helpful discussions and comments. The work of SK is supported by the National Research Foundation of Korea (NRF) Grant RS-2024-00405629 and RS-2023-NR076601. The work of SS is supported by DE-SC0010008.

\appendix

\section{Tools: supersymmetric observables}
\label{app: A-model}
In this appendix, we review some 3d supersymmetric observables, with particular interest in 3d $\CN=2$ abelian Chern-Simons matter theories.

\subsection{Twisted partition function} 
The 3d A-model method is developed in \cite{Closset:2018ghr} to compute the half-BPS partition functions of 3d $\CN=2$ gauge theories on any compact Seifert 3-manofold. See \cite{Closset:2019hyt} for a review, and \cite{Closset:2023vos,Closset:2023jiq,Closset:2023bdr,Closset:2023izb,Closset:2024sle,Closset:2025lqt} for recent applications. The idea is to use the topologically A-twisted 2d $\CN=(2,2)$ theory together with the Bethe/gauge correspondence \cite{Nekrasov:2009uh,Nekrasov:2009ui}.

The twisted partition function $Z_{\CM_{g,p}}$ on a degree $p$ bundle over a genus $g$ Riemann surface $S^1 \overset{p}{\to} \CM_{g,p} \to \S_g$ can be evaluated as~\cite{Closset:2018ghr}
\begin{align}
    Z_{\CM_{g,p}} = \sum_{u^{(\a)} \in \CS} \big(\CH_\a\big)^{g-1} \big(\CF_\a\big)^p
    \,,
    \label{eq: ptf Mgp}
\end{align}
where $\CH$ and $\CF$ are {\it handle gluing} and {\it fibering} operators respectively that can be computed from 3d $\CN=2$ gauge theory description. The sum is over the {\it Bethe vacua} $ u^{(\a)} \in  \CS$ labeled by $\a$. Let us explain these ingredients in what follows.

\medskip\noindent{\bf Bethe vacua}.
For a 3d $\CN=2$ gauge theory with gauge group $G$ and some matters, the 3d twisted superpotential $\CW(u,\n)$ and the effective dilaton $\Omega(u,\n)$ can be established and they characterize the Coulomb branch low-energy dynamics with the gauge and flavor symmetry parameters
\begin{align}
    u = (u_1,\cdots,u_{\rank(G)}) \, , \quad
    \n = (\n_1,\cdots,\n_{\rank(F)})\,,
\end{align}
where $F$ is flavor symmetry group. The gauge and flavor symmetry flux operators are defined respectively as
\begin{align}
    \Pi_j(u,\n) :=
    \exp(
    2\pi i \frac{\partial \CW(u,\n) }{\partial u_j} ) \, , \quad
    \Pi_a^f(u,\n) :=
    \exp(
    2\pi i \frac{\partial \CW(u,\n)}{\partial \n_a}
    )
    \,.
\end{align}
Then, the Bethe vacua are given as a set 
\begin{align}
    \CS = \left\{ u^{(\a)} \,\middle| \,
    \Pi_j(u,\n) = 1 \, , \ 
    j=1,\cdots,\rank(G) \, , \
    w(u) \neq u \, , \
    \forall w\in W_G
    \middle\} \right/ W_G \, .
\end{align}
where $W_G$ is Weyl group of $G$. Once we replace the gauge variables into holomorphic ones
\begin{align}
    z_i := e^{2\pi i\, u_i} \, , \quad
    y_a := e^{2\pi i\, \n_a}\, ,
\end{align}
the equations become polynomials in $z_i$ so that the number of Bethe vacua $|\CS|$ is finite, and hence so is the sum in \eqref{eq: ptf Mgp}.

\medskip\noindent{\bf Handle gluing and fibering operators}.
The handle gluing and fibering operators are evaluated as
\begin{align}
    \begin{aligned}
        \CH(u,\n) &= e^{2\pi i \Omega(u,\n)} 
        \det(
        \frac{\partial^2 \CW(u,\n)}{\partial u_i \partial u_j}
        ) \, , \\
        \CF(u,\n) &= 
        \exp(
        2\pi i \left(
            1
            - \sum_{i=1}^{\text{Rank}(G)}\!\! u_i \, \frac{\partial }{\partial u_i}
            - \sum_{a=1}^{\text{Rank}(F)}\!\! \n_a \, \frac{\partial }{\partial \n_a}
        \right)\CW (u,\n)
        )\,.
    \end{aligned}
    \label{eq: H and F}
\end{align}
so that $\CH_\a$ and $\CF_\a$ in \eqref{eq: ptf Mgp} are the values at the $\a$-th Bethe vacuum
\begin{align}
    \CH_\a := \CH(u^{(\a)},\n) \, , \quad
    \CF_\a := \CF(u^{(\a)},\n)\,.
\end{align}

\subsubsection{3d \texorpdfstring{$\CN=2$}{N=2} Abelian Chern-Simons matter theory}
Let us apply the A-model method for an ACSM theory of gauge group $U(1)^r$ and $N$ chiral multiplets $\Phi_{I=1,\cdots,N}$ with $r\times r$ CS level matrix $K$ and $r\times N$ charge matrix $C$
\begin{align}
    K = \left(
    \begin{array}{ccc}
         K_{11}  & \cdots & K_{1r} \\
         \vdots & \ddots & \vdots  \\
         K_{1r} & \cdots  & K_{rr}
    \end{array}
    \right) \, , \quad
    C = \left(
    \begin{array}{cccc}
         C_{11} &  \cdots &  & C_{1N} \\
         \vdots & \ddots & & \vdots \\
         C_{r1} & \cdots &  & C_{rN}
    \end{array}
    \right)\, ,
\end{align}
where $K$ is symmetric, and $C_{iJ}$ is the gauge charge of the $J$-th chiral multiplet under the $i$-th $U(1)$ factor. This class of theories has attracted considerable interest as the best way of constructing UV gauge theory description of non-unitary TQFTs~\cite{Gang:2023rei,Ferrari:2023fez,Baek:2024tuo,Creutzig:2024ljv,Gang:2024loa,Gang:2025ykf,Gang:2026iem}; see also~\cite{Gang:2022kpe,Gang:2024tlp,Jeong:2025xid,Jeong:2026dzz} for the construction from the S-fold theories. The twisted superpotential and the effective dilaton of this theory read
\begin{align}
    &\CW(u,\n) =
    \frac{1}{2}\sum_{i,j=1}^r K_{ij} u_i u_j
    +\frac{1}{2} \sum_{i=1}^r \big((1+2\n_R)K_{ii} + 2\n_i\big) u_i + \frac{1}{(2\pi i)^2}\sum_{I=1}^N \Li_2 \big( e^{2\pi i u\cdot C_I} \big)
    \,,
    \nonumber\\
    &\Omega(u) = \frac{1}{2\pi i} \sum_{I=1}^N \log \big(
    1-e^{2\pi i u\cdot C_I}
    \big) + \frac{1}{2}k_{RR}\,,
\end{align}
where $\n_R \in \frac{(p\,\text{mod}\,2) + 1}{2}\mathbb{Z}$ and $k_{RR}$ are the mass parameter and CS level of $U(1)_R$ symmetry, respectively.~\footnote{We take $\n_R = 1$ when we extract the modular data in the main text.} By assuming $r \geq N$ and $\text{rank}(C) = N$, all the flavor symmetries, parametrized by $\n_i$, come from the $U(1)_{T_i}$ topological symmetries. Then, the Bethe vacuum equations read
\begin{align}
    (-1)^{(1+2\n_R)K_{ii}} y_i
    \prod_{j=1}^r z_j^{K_{ij}} 
    =\prod_{I=1}^N \big(1-z^{C_I}\big)^{C_{iI}} \quad \text{for} \quad i=1,\cdots,r
    \,,
    \label{eq: ACSM Bethe}
\end{align}
with a short-hand notation $z^{C_I} := \prod_{i=1}^r z_i^{C_{iI}}$. The handle gluing and fibering operators are computed as
\begin{align}
    &\CH(u) =(-1)^{k_{RR}} \prod_{I=1}^N \Big( 1- z^{C_I} \Big)
    \det_{i,j}
    \Bigg(
    K_{ij} + \sum_{I=1}^N C_{iI} C_{jI} \frac{z^{C_I}}{1-z^{C_I}}
    \Bigg) \, ,
    \nonumber\\
    &\CF(u,\n) = 
    \prod_{I=1}^N
    \big(1-z^{C_I}\big)^{u\cdot C_I}
    \exp(
    \frac{1}{2\pi i}
    \sum_{I=1}^N
     \Li_2\big( z^{C_I} \big)
     -\pi i\sum_{i,j=1}^r K_{ij}u_i u_j - 
     2\pi i \sum_{i=1}^r \n_i u_i
    ) \, .
    \label{eq: ACSM HF}
\end{align}
The partition function with a Wilson loop insertion $W_Q$ of charge $Q$ reads
\begin{align}
    \ang{W_Q}_{\CM_{g,p}}
    =
    \sum_{\a} 
    e^{-2\pi i Q \cdot u^{(\a)}}
    (\CH_\a)^{g-1}
    (\CF_\a)^p
    \,.
\end{align}

\medskip
\noindent{\bf Monopole superpotential and mixing}.
The ACSM theories we consider include gauge invariant half-BPS dressed monopole operators
\begin{align}
    V_{(d,\mathfrak{m})}
    :=
    \Big(\prod_{I=1}^{N} \phi_I^{d_I} \Big)
    V_{\mathfrak{m}}
    \,,
    \label{eq: half BPS monopole}
\end{align}
where $d$ collects the dressing numbers of scalar field $\phi_I$ in the $I$-th chiral multiplet $\Phi_I$, while $V_\mathfrak{m}$ denotes a bare monopole operator with magnetic flux $\mathfrak{m}$. The R-charge of the monopole operator with mixing of the topological symmetries parametrized by $\m$ is computed as
\begin{align}
    R(V_{(d,\mathfrak{m})})
    =
    \sum_{I| \mathfrak{m}\cdot C_I >0}
    \!\!
    \mathfrak{m} \cdot C_I
    +
    \m\cdot \mathfrak{m}
    \,,
\end{align}
so that each monopole superpotential term gives R-charge 2 constraint $R(V_{(d,\mathfrak{m})})=2$. Thus, the superpotential terms solve $\m$ in terms of unbroken combinations of the topological symmetries. This mixing fixes the flavor mass parameter to $\n = \m \n_R$ and shifts the handle gluing and fibering operators by the flavor flux operator, resulting in the partition function
\begin{align}
    Z_{\CM_{g,p}}
    =
    \sum_{ u^{(\a)}\in \CS }
    \Big[
    \underbrace{
    \CH(u^{(\a)},\m \n_R )
    \Pi^f(u^{(\a)},\m\n_R)^{\m}
    }_{\tilde{\CH}_\a}
    \Big]^{g-1}
    \Big[
    \underbrace{
    \CF(u^{(\a)},\m \n_R )
    \big(\Pi^f(u^{(\a)},\m\n_R)^{\m}\big)^{\n_R}
    }_{\tilde{\CF}_\a}
    \Big]^{p}
    \,,
\end{align}
where $\Pi^f(u^{(\a)},\m\n_R)^\m := \prod_{a}\Pi_a^f(u^{(\a)},\m\n_R)^{\m_a}$. If this ACSM theory flows to a semisimple TQFT, the partition function will be written in terms of the modular matrices as
\begin{align}
    Z_{\CM_{g,p}}^{\rm TQFT} 
    =
    \sum_{\a}
    (S_{0\a})^{2-2g} (T_{\a\a})^{-p}
    \,,
\end{align}
which suggests a map
\begin{align}
    \tilde{\CH}_\a = (S_{0\a})^{-2}
    \;,\;\;
    \tilde{\CF}_\a = (T_{\a\a})^{-1}
    \,,
    \label{eq: HSFT}
\end{align}
up to some normalization factors. We use this relation to compute the associated modular data of the IR TQFT phase from the 3d $\CN=2$ UV gauge theory. For notational simplicity, we simply denote the shifted quantities ($\tilde{\CH}_\a$,$\tilde{\CF}_\a$) by ($\CH_\a$,$\CF_\a$) in the main text.

\subsection{Superconformal index}
The 3d $\CN=2$ superconformal index counts gauge invariant local BPS operators~\cite{Kim:2009wb,Imamura:2011su}
\begin{align}
    \CI_{S^2 \times S^1}(y;q)
    =
    \Tr (-1)^R q^{\frac{R}{2}+j_3} y^f
    \,,
\end{align}
where the trace is over the radially quantized Hilbert space on $S^2$, $R$ is the conformal $R$-charge, $j_3$ is the Cartan of the $SO(3)$ isometry of $S^2$, and $y$ denotes flavor symmery fugacity. 

\medskip\noindent{\bf ACSM theory}. 
For the ACSM theory, we use the localization formula~\cite{Imamura:2011su,Kapustin:2011jm}
\begin{align}
    \CI_{S^2\times S^1}(y;q)
    =
    \sum_{\mathfrak{m}\in \mathbb{Z}^r}
    \oint
    \prod_{i=1}^r
    \frac{dz_i}{2\pi i z_i}
    z_i^{\sum_{j=1}^r K_{ij}\mathfrak{m}_j}
    y_i^{\mathfrak{m}_i}
    \prod_{I=1}^N
    \CI_\D
    ( \mathfrak{m} \cdot C_I, z^{C_I} ;q)
    \,,
    \label{eq: SCI formula}
\end{align}
with the chiral contribution
\begin{align}
    \CI_\D(\mathfrak{m},z;q)
    =
    \prod_{l=0}^\infty
    \frac{1-q^{l-\frac{\mathfrak{m}}{2}+1}z^{-1}}{1-q^{l-\frac{\mathfrak{m}}{2}} z  }
    \,.
\end{align}
Suppose the mixing parameter $\m$ is given in terms of $n$ remaining combinations of UV topological symmetries and the fixed point $\m_0$ as
\begin{align}
    \m = \m_0 + \sum_{j=1}^n a^{(j)} \n^{(j)}
    \,,
\end{align}
where $a^{(j)} \in \mathbb{Z}^r$ describes the $j$-th remaining combination parametrized by $\n^{(j)}\in\mathbb{R}$. Then, the flavor symmetry fugacities $y_i$ are rewritten in terms of $\eta_j$ for the remaining symmetries
\begin{align}
    y_i = (-q^{\frac{1}{2}})^{(\m_0)_i} \prod_{j=1}^n \eta_j^{a^{(j)}_i}
    \,.
\end{align}

\medskip\noindent{\bf Wilson loop}. The BPS Wilson loop operator $W_Q$ of charge $Q=(Q_1,\cdots,Q_r)$ wrapping the $S^1$ direction on the superconformal index background can only be closed when it is located on the north/south pole of $S^2$~\cite{Dimofte:2011py,Gang:2009qdj}. We denote it by
\begin{align}
    W_Q^{\pm}(\mathfrak{m},z)
    :=
    \prod_{i=1}^r
    \big(
    q^{\frac{\mathfrak{m_i}}{2}}
    z_i^{\pm 1}
    \big)^{Q_i}
    \label{eq: index Wilson}
\end{align}
and its insertion in the formula \eqref{eq: SCI formula}, then, simply reads
\begin{align}
    \ang{W_Q^\pm}_{\rm sci}
    :=
    \sum_{\mathfrak{m}\in \mathbb{Z}^r}
    \oint
    \prod_{i=1}^r
    \frac{dz_i}{2\pi i z_i}
    z_i^{\sum_{j=1}^r K_{ij}\mathfrak{m}_j \pm Q_i}
    \big(
    y_i q^{\frac{Q_i}{2}}
    \big)^{\mathfrak{m}_i}
    \prod_{I=1}^N
    \CI_\D
    ( \mathfrak{m} \cdot C_I, z^{C_I} ;q)
    \,.
\end{align}
The generalization to multiple Wilson loop insertions $\ang{W_{Q}^{\s} W_{Q'}^{\s'} \cdots}_{\rm sci} $ is straightforward. This provides a non-trivial necessary condition for $W_Q$ to be mapped to a simple object in the IR TQFT phase~\cite{Gang:2024loa}
\begin{align}
    \ang{W_Q^\pm}_{\rm sci} = 0
    \;,\;\;
    \ang{W_Q^+W_Q^-}_{\rm sci} = 1
    \,,
    \label{eq: SCI simple line condition}
\end{align}
which are motivated by the fusion rules of the simple objects.

Our goal in the main text is to identify simple lines in terms of UV Wilson lines whose existence is not guaranteed. If we could not identify a Wilson line, which is a monomial as in \eqref{eq: index Wilson}, we try with multinomial
\begin{align}
    W_{\mathfrak{Q},\mathfrak{R},\e}^{\pm}(\mathfrak{m},z)
    :=
    \sum_{i} 
    \e_i
    (-q^{\pm \frac{1}{2}})^{R^{(i)}}
    \prod_{j=1}^{r}
    (
    q^{\frac{\mathfrak{m}_j}{2}}
    z_j^{\pm 1}
    )^{Q_j^{(i)}}
\end{align}
where $\mathfrak{Q} = \{ Q^{(1)},Q^{(2)},\cdots \}$ and $\mathfrak{R} = \{ R^{(1)},R^{(2)},\cdots \}$ are ordered set of electric and R-charge of each Wilson loop term respectively, while $\e_i \in \{+1,-1\}$ is a possible sign factor. The R-charges $R^{(i)}$'s and signs $\e_i$ are only relatively defined. We straightforwardly generalize \eqref{eq: SCI simple line condition} as
\begin{align}
    \ang{W_{\mathfrak{Q},\mathfrak{R}}^\pm}_{\rm sci} = 0
    \;,\;\;
    \ang{W_{\mathfrak{Q},\mathfrak{R}}^+W_{\mathfrak{Q},\mathfrak{R}}^-}_{\rm sci} = 1
    \,,
\end{align}
to identify a multinomial loop operator that maps to a simple line in the IR TQFT. Its A-model version becomes
\begin{align}
    \ang{W_{\mathfrak{Q},\e}}_{\CM_{g,p}}
    =
    \sum_{\a}
    W_{\mathfrak{Q},\e}(u^{(\a)})
    (\CH_\a)^{g-1} (\CF_\a)^p
    \;;
    \;\;
    W_{\mathfrak{Q},\e}(u)
    :=
    \sum_{j} \e_j e^{-2\pi i Q^{(j)} \cdot u }
    \,,
\end{align}
which does not depend on $\mathfrak{R}$.

%%%%%%%%%%%%%%%%%%%%%%%%%%%%%%%%%%%%%%%%%%%%%%%%%%%%%%%

\section{Supplementary computations on \texorpdfstring{$A_3$}{A3} theory} \label{app: A3 monodromy action}
We provide the monodromy action on the half-BPS line defects of the $A_3$ theory and the trace computation. The monodromy operator in the minimal chamber can be simplified
\begin{align}
    \Phi = (1)(3)(2)(-1)(-3)(-2)
    =
    (x) e^{p+m} e^{p-m} e^x (-p-m)(-p+m)
    \,,
\end{align}
where we defined $\g_p := \frac{\g_1 + \g_3}{2}$, $\g_x := \g_2$, and $\g_m := \frac{\g_1-\g_3}{2}$, with $\ang{\g_p,\g_x}=1$. Here we first note that for any such unit Dirac pairing variables $p$ and $x$ satisfy the below
\begin{align}
    (p)[x] &= [x](p-i b) =
    [x]\big\{ {\bf 1} + q^{1/2} [p] \big\} (p)
    =
    \big\{ [x] + [x+p] \big\} (p)
    \,,
    \nonumber\\
    (x)[p] &= [p](x+i b) =
    [p]\big\{ {\bf 1} + q^{-1/2} [x] \big\}^{-1} (x)
    \,.
    \label{eq: line shifts}
\end{align}

\subsection{Monodromy action}
We verify the monodromy action \eqref{eq: A3 monodromy action} by using the properties in \eqref{eq: line shifts}. We can express the line operators in terms of the abbreviations \eqref{eq: abbreviation}:
\begin{align}
    &F(\CA_1) = [p]
    \,,
    \nonumber\\
    &F(\CA_2) = [-p-x]+[-p]
    \,,
    \nonumber\\
    &F(\CA_3) = [-p] + C [x] +[p+x] + [-p+x]
    \,,
    \nonumber\\
    &F(\CB_1) = [-x]
    \,,
    \nonumber\\
    &F(\CB_2) = [x] + C\{[-p]+[-p+x] \} 
    +\{q^{\frac{1}{2}}+q^{-\frac{1}{2}}\}[-2p] + [-2p+x] + [-2p-x]
    \,,
    \nonumber\\
    &F(\CB_3) = [x] + C[p+x] + [2p+x]
    \,,
    \nonumber\\
    &C : = F(\CC)  = [m]+[-m]
    \,,
\end{align}
then, we find:
\begin{itemize}
    \item $\Phi F(\CA_1) = F(\CA_2) \Phi$
    \begin{align}
        \Phi  F(\CA_1) & = 
        (x) e^{p+m} e^{p-m} e^x (-p-m)(-p+m) [p]
        \nonumber\\
        &=
        (x) [-p-x] e^{p+m} e^{p-m} e^x (-p-m)(-p+m)
        \nonumber\\
        &= \{ [-p-x] + [-p] \} \Phi
        \nonumber\\
        &= F(\CA_2) \Phi
        \,.
    \end{align}

    \item $ F(\CA_1) \Phi = \Phi F(\CA_3)$
    \begin{align}
        F(\CA_1) \Phi 
        & = 
        [p] (x) e^{p+m} e^{p-m} e^x (-p-m)(-p+m)
        \nonumber\\
        & = 
        (x) \big\{ [p] + [p+x] \big\}  e^{p+m} e^{p-m} e^x (-p-m)(-p+m)
        \nonumber\\
        & = 
        (x) e^{p+m} e^{p-m} e^x \big\{ [p+x] + [-p] \big\} (-p-m)(-p+m)
        \nonumber\\
        & = 
        \Phi
        \big\{ [x-m] + [x+m] + [-p+x] + [p+x] + [-p] \big\}
        \nonumber\\
        &= \Phi F(\CA_3)
        \,.
    \end{align}

    \item $ \Phi F(\CA_2) = F(\CA_3) \Phi $
    \begin{align}
        \Phi F(\CA_2)
        & = 
        (x) e^{p+m} e^{p-m} e^x (-p-m)(-p+m)
        \{ [-p-x] + [-p] \}
        \nonumber\\
        & = 
        (x) e^{p+m} e^{p-m} e^x
        \big\{ [-p-x] + [-3p-x] 
        \nonumber\\
        &\qquad\qquad + C[-2p-x] + [-p] \big\}
        (-p-m)(-p+m)
        \nonumber\\
        & = 
        (x)
        \big\{ 
        [-p] + [p+2x] + C[x] + [p+x] 
        \big\}
        e^{p+m} e^{p-m} e^x(-p-m)(-p+m)
        \nonumber\\
        & = 
        \Big\{ 
        [-p] + [-p+x] + [p+2x] 
        \{ 1+q^{-1/2} [x] \}^{-1} 
        \nonumber\\
        &\qquad\qquad\qquad+C[x]
        +[p+x] \{ 1+q^{-1/2} [x] \}^{-1} 
        \Big\}
        \Phi
        \nonumber\\
        &= \big\{
        [-p] + [-p+x] + C[x] + [p+x]
        \big\}\Phi
        \nonumber\\
        &= F(\CA_3) \Phi
        \,.
    \end{align}

    \item $F(\CB_1) \Phi = \Phi  F(\CB_3) $
    \begin{align}
        F(\CB_1) \Phi
        & = 
        [-x]
        (x) e^{p+m} e^{p-m} e^x (-p-m)(-p+m)
        \nonumber\\
        & = 
        (x) e^{p+m} e^{p-m} e^x 
        [2p+x]
        (-p-m)(-p+m)
        \nonumber\\
        & = 
        (x) e^{p+m} e^{p-m} e^x (-p-m)
        \big\{
        [2p+x] + [p+x-m]
        \big\}
        (-p+m)
        \nonumber\\
        & = 
        \Phi
        \big\{
        [p+x+m] + [p+x-m] + [x] + [2p+x]
        \big\}
        \nonumber\\
        & = \Phi F(\CB_3)
        \,.
    \end{align}

    \item $ \Phi F(\CB_1) = F(\CB_2) \Phi  $
    \begin{align}
        \Phi F(\CB_1)
        & = 
        (x) e^{p+m} e^{p-m} e^x (-p-m)(-p+m)
        [-x]
        \nonumber\\
        & =
        (x) e^{p+m} e^{p-m} e^x 
        \{
        [-x] + C[-p-x]
        + [-2p-x]
        \}
        (-p-m)(-p+m)
        \nonumber\\
        & =
        (x) 
        \{
        [-2p-x] + [x] + C[-p]
        \}
         e^{p+m} e^{p-m} e^x
        (-p-m)(-p+m)
        \nonumber\\
        & =
        \{
        [-2p-x] + (q^{\frac{1}{2}}+q^{-\frac{1}{2}}) [-2p] + [-2p+x]
        +[x] + C[-p] + C[-p+x]
        \}
        \Phi
        \nonumber\\
        & = F(\CB_2) \Phi
        \,.
    \end{align}

    \item $ F(\CB_3) \Phi = \Phi F(\CB_2)$
    \begin{align}
        F(\CB_3) \Phi 
        & = 
        \{ C[p+x] + [x] + [2p+x] \}
        (x) e^{p+m} e^{p-m} e^x (-p-m)(-p+m)
        \nonumber\\
        & =
        (x) e^{p+m} e^{p-m} e^x
        \Big\{
        [-3p-x] + [-p]
        + [-2p-x]
        \nonumber\\
        &\qquad 
        \{[-4p-x] + [x] + \{ q^{\frac{1}{2}} + q^{-\frac{1}{2}} \} [-2p] \}
        \Big\}
        (-p-m)(-p+m)
        \nonumber\\
        &= \Phi
        \Big\{
        C[-p]
        \{q^{\frac{1}{2}} + q^{-\frac{1}{2}}\} [-2p]
        + [x]
        + C [-p+x]
        + [-2p+x]
        \nonumber\\
        &\qquad\quad +
        \big\{ 1 + q^{-\frac{1}{2}} [-p-m] \big\}^{-1}
        \big\{ 1 + q^{-\frac{1}{2}} [-p+m] \big\}^{-1}
        \nonumber\\
        &\qquad\qquad \times 
        \big\{
        [-2p-x]
        +C[-3p-x]
        [-4p-x]
        \big\}
        \Big\}
        \nonumber\\
        & = \Phi
        \Big\{
        [x] + [-2p+x] + [-2p-x] + C[-p] + C[-p+x] + \{q^{\frac{1}{2}}+q^{-\frac{1}{2}}\}[-2p]
        \Big\}
        \nonumber\\
        &= \Phi F(\CB_2)
        \,.
    \end{align}
    
\end{itemize}
These explicitly demonstrates the monodromy action \eqref{eq: A3 monodromy action} that implies the degeneracy independence \eqref{eq: A3 deg indep} in the trace.

\subsection{Evaluation of the trace with line defect insertion}
Here we show the ellipsoid partition function with the line defect insertion \eqref{eq: A3 integral} by evaluating the trace. We first simplify the monodromy operator as
\begin{align}
    \Phi &= (1)(3)(2)(-1)(-3)(-2)
    \nonumber\\
    &=(2)e^3 (2) e^1 e^2 (-3)
    \nonumber\\
    &= (x) e^{p-m} (x) e^{p+m} e^x (-p+m)
    \,.
\end{align}
Then, we further simplify the traces as
\begin{align}
    \ang{\CA}_b
    &=
    \Tr(F[\CA_1])
    \nonumber\\
    &= \Tr \big(
    [p] (x) e^{p-m} (x) e^{p+m} e^x (-p+m)
    \big)
    \nonumber\\
    &= \Tr \big(
    [p] (-p+m) (x) (p+x-m) e^{p-m} e^{p+m} e^x 
    \big)
    \nonumber\\
    &= \Tr \big(
    [p] (e^{x})^{-1} (p-m) e^x (-p+m) e^{p-m} e^{p+m} e^x
    \big)
    \,,
    \\
    \ang{\CB}_b
    &=
    \Tr(F[\CB_1])
    \nonumber\\
    &= \Tr \big(
    [-x] (x) e^{p-m} (x) e^{p+m} e^x (-p+m)
    \big)
    \nonumber\\
    &= \Tr \big(
    [-x] (-p+m-ib)(x)(p+x-m) e^{p-m} e^{p+m} e^x
    \big)
    \nonumber\\
    &= \Tr \big(
    [-x] \big\{
    1 + e^{-\pi i b^2}[-p+m]
    \big\}
    (e^{x})^{-1} (p-m) e^x (-p+m) e^{p-m} e^{p+m} e^x
    \big)
    \,,
\end{align}
and by inserting three pairs of completeness and doing gaussian integrals, we find
\begin{align}
    \ang{\CA}_b 
    &= 
    i^{\frac{1}{2}}e_b^6
    \int d p_1 d p_2 d p_3
    e^{2\pi b p_1} e^{\pi i ( 2m^2 - 2 p_1 p_2 + 2 p_1 p_3 + 2 p_2 p_3  )}
    \Phi_b(p_2 - m)\Phi_b(-p_3 + m)
    \nonumber\\
    &= i^{\frac{1}{2}}e_b^6
    \int dp_3 e^{\pi i (2m^2 + 2p_3^2 -2ib p_3)}
    \Phi_b(p_3 - m -ib)\Phi_b(-p_3+m)
    \nonumber\\
    &
    \overset{p_3 \to \s}{=} 
    i^{\frac{1}{2}}e_b^6
    \int d\s 
    e^{\pi i ( 2m^2 + 2\s^2 - 2ib \s )}
    \big(
    1 + e^{-\pi i b^2} e^{2\pi b \s} e^{-2\pi b m}
    \big) 
    \underbrace{\Phi_b(\s - m) \Phi_b(-\s+m)}_{=e_b^2 e^{\pi i (\s-m)^2}}
    \nonumber\\
    &=
    i^{\frac{1}{2}}e_b^8
    \int d\s 
    \big(
    e^{2\pi b \s }
    +
    e^{-\pi i b^2} e^{4\pi b \s} e^{-2\pi b m}
    \big)
    e^{\pi i (3\s^2 - 2m\s + 3m^2)}
    \\
    \ang{\CB}_b
    &=
    i^{\frac{1}{2}} e_b^6
    \int dp_1 dp_2 dx_2
    e^{-2\pi b x_2} e^{\pi i (2m^2 - p_1^2 + 2p_1 p_2 + p_2^2 + 2p_1 x_2 - 2p_2 x_2)}
    \Phi_b(p_1-m)\Phi_b(-p_2+m)
    \nonumber\\
    &\quad+i^{\frac{1}{2}} e_b^6 
    \int dp_1 dp_2 dp_3
    e^{2\pi b m} e^{-2\pi b p_3} 
    e^{\pi i (2m^2 - 2p_1 p_2 + 2 p_1 p_3 + 2p_2 p_3)}
    \Phi_b(p_2-m) \Phi_b(-p_3+m)
    \nonumber\\
    &=i^{\frac{1}{2}} e_b^8 e^{-\pi i b^2}
    \int dp_1 e^{\pi i (2m^2 + 2p_1^2 + 4i b p_1)}
    \big(1 + e^{-\pi i b^2}e^{2\pi b(-p_1+m)} \big) e^{\pi i (p_1-m)^2}
    \nonumber\\
    &\qquad+i^{\frac{1}{2}} e_b^8
    \int dp_2 e^{2\pi b m } e^{-2\pi b p_2} e^{\pi i (2m^2 + 2p_2^2)} e^{\pi i(p_2-m)^2}
    \nonumber\\
    &\overset{p_1,p_2\to\s}{=}
    i^{\frac{1}{2}} e_b^8
    \int d\s
    \big( 
    e^{2\pi b m} e^{-2\pi b \s}
    \!+\!
    e^{-\pi i b^2} e^{-4\pi b \s}
    \!+\!
    e^{-2\pi i b^2} e^{2\pi b m} e^{-6\pi b \s}
    \big)
    e^{\pi i (3\s^2 - 2 m \s + 3 m^2)}
    \,.
\end{align}

\bibliographystyle{JHEP}
\bibliography{ref}
\end{document}